\documentclass{aa}  

\usepackage{graphicx}

\usepackage{txfonts}
 \usepackage{hyperref}
 \hypersetup{
     colorlinks=true,
     linkcolor=blue,
     filecolor=blue,
     citecolor = blue,      
     urlcolor=cyan,
     }
\usepackage{subfigure}
\usepackage{longtable}
\usepackage{units}
\usepackage{txfonts}
\usepackage{color}

\begin{document}

   \title{Self-scattering of non-spherical dust grains}
   \subtitle{The limitations of perfect compact spheres}   
   \titlerunning{Self-scattering of non-spherical dust grains}

\author{Florian Kirchschlager\inst{1} \and Gesa H.-M. Bertrang\inst{2}}
 \institute{Department of Physics and Astronomy, University College London, Gower Street, London WC1E 6BT, United Kingdom\\
 \email{f.kirchschlager@ucl.ac.uk}
 \and
 Max Planck Institute for Astronomy, K\"onigstuhl 17, 69117 Heidelberg, Germany\\
  \email{bertrang@mpia.de}
 }

 \date{Received 12 March 2020 / Accepted 28 April 2020}

 
  \abstract
   {The understanding of (sub-)millimetre~polarisation has made a leap forward since high-resolution imaging with the Atacama Large (sub-)Mm Array (ALMA) came available.
    Amongst other effects, self-scattering (i.e., scattering of thermal dust emission on other grains) is thought to be the origin of millimetre~polarisation. This opens the first window to a direct measurement of dust grain sizes in regions of optically thick continuum emission as it can be found in protoplanetary disks and star-forming regions. However, the newly derived values of grain sizes are usually around ${\sim}\unit[100]{\mu m}$ and thus one order of magnitude smaller than those obtained from more indirect measurements as well as those expected from theory (${\sim}\unit[1]{mm}$).} 
   {We see the origin of this contradiction in the applied dust model of today's self-scattering simulations: a perfect compact sphere. The aim of this study is to test our hypothesis by investigating the impact of non-spherical grain shapes on the self-scattering signal.}
    {We apply discrete dipole approximation simulations to investigate the influence of the grain shape on self-scattering polarisation in three scenarios: an unpolarised and polarised incoming wave under a fixed as well as a varying incident polarisation angle.}
   {We find significant deviations of the resulting self-scattering polarisation when comparing non-spherical to spherical grains. In particular, tremendous deviations are found for the polarisation signal of grains when observed outside the Rayleigh regime, i.e. for $\unit[{>}100]{\mu m}$ size grains observed at $\unit[870]{\mu m}$ wavelength. Self-scattering by oblate grains produces higher polarisation degrees compared to spheres which challenges the interpretation of the origin of observed millimetre polarisation. A (nearly) perfect alignment of the non-spherical grains is required to account for the observed millimetre polarisation in protoplanetary disks. Furthermore, we find conditions under which the emerging scattering polarisation of non-spherical grains is flipped in orientation by 90$^{\circ}$. }
   {These results show clearly that the perfect compact sphere is an oversimplified model which reached its limit. Our findings point towards a necessary re-evaluation of the dust grain sizes derived from (sub-)mm polarisation.}

\keywords{polarisation -- scattering -- protoplanetary disks -- stars: circumstellar matter -- stars: pre-main sequence  -- techniques: polarimetric}
   \maketitle
 

\section{Introduction}
In recent years, the field of polarisation at millimetre~wavelengths went through remarkable development. Polarisation became a powerful and versatile tool to probe not only magnetic fields (\citealt{Lazarian2007, Bertrang2017a,Bertrang2017b}) but also grain sizes (\citealt{Kataoka2015}), radiation fields \citep{Lazarian2007, Tazaki2017}, and grain porosity (\citealt{Kirchschlager2019}). Particularly, the ability to measure dust grain sizes pushed the protoplanetary disk community's interest in self-scattering. The number of cutting-edge polarisation observations, obtained with the Atacama Large (sub-)Mm Array (ALMA), is continuously growing for both protoplanetary disks \citep[e.g.,][]{Kataoka2017, Stephens2017, Dent2019, Harrison2019} as well as star-forming regions \citep[e.g.,][]{Bacciotti2018, Sadavoy2019}. Previous to the discovery of self-scattering, dust grain sizes were deduced only indirectly by relying on various assumptions on optical dust properties \citep[{\it ``spectral index method''}; e.g.,][]{Miyake1993, Carrasco2019}. However, the grain sizes which are inferred from self-scattering are challenging the understanding of grain growth: the newly derived grain sizes (${\sim}\unit[100]{\mu m}$) are one order of magnitude smaller than those derived from the spectral index (\citealt{Beckwith1990}) and those expected from theory (${\sim}\unit[1]{mm}$). In this paper, we tackle our hypothesis about the cause of this contradiction: the oversimplified dust model of a perfect compact sphere which is applied in current self-scattering models (e.g.,~\citealt{Kataoka2017, Dent2019}). The result of this study impacts the interpretation of basic dust properties and as such, can be applied to various astrophysical environments such as protoplanetary disks, star-forming regions or the interstellar medium.

 In general, continuum radiation is intrinsically polarised as grains are elongated (e.g., \citealt{Lazarian2007, Bertrang2017a,Bertrang2017b,Kirchschlager2019,Guillet2020}). In the specific case of optically thick continuum emission, the intrinsically polarised dust emission gets scattered at other dust grains (self-scattering). In this paper, we investigate the impact of the shape of perfectly aligned grains on the self-scattering signal, including the intrinsic polarisation of the wave before the scattering event.
 
 In Section~\ref{201} we describe our non-spherical dust grain model and the method to calculate their optical properties. We present the results for self-scattering without intrinsic polarisation on oblate dust grains in Section~\ref{301} and for the self-scattering of intrinsically polarised radiation in Section~\ref{401}. We discuss our findings in Section~\ref{501} and conclude with a summary in Section~\ref{601}.


\section{Dust model and methods}
\label{201}
In our study we considered oblate dust grains with short semi-axis $b$ and long semi-axis $c$. We varied the axis ratio $\nicefrac{c}{b}$ in the set $\left\lbrace 1.0,1.1,1.3,1.5,1.7,2.0\right\rbrace$, where $\nicefrac{c}{b}=1.0$ represents a spherical grain. We also varied the effective radius of the oblate grains which is the radius of a volume-equivalent solid sphere and that is given as $a_\text{eff}=\left(bc^2\right)^{1/3}$. The particles are compact so the porosity is $\mathcal{P}=0.0$.

To calculate the optical properties of irregular shaped particles we used the code \textsc{DDSCAT}\footnote{\href{https://www.astro.princeton.edu/~draine/DDSCAT.7.3.html}{https://www.astro.princeton.edu/~draine/DDSCAT.7.3.html}} (version 7.3; \citealt{Draine1994,Draine2013}) which is based on the theory of discrete dipole approximation (DDA; \citealt{PurcellPenny1973}). The three-dimensional particle shape is replaced by a corresponding spatial distribution of $N$ discrete dipoles on a cubic grid and the optical properties are then calculated for this dipole distribution. 
The \textsc{DDSCAT} is well tested and is applicable for most particle shapes and structures but is limited by an upper value for the ratio of grain size $a_\text{eff}$ to wavelength $\lambda$ (see e.g.,~\citealt{DraineGood, Draine1994, Kirchschlager2013}). In our study, we used $N= 145850$ dipoles which corresponds to an upper limit of $a_\text{eff}/\lambda\lesssim 1.5$.

\begin{figure}
 \fbox{\includegraphics[trim=3cm 5.5cm 0cm 4.2cm,  clip=true,page=1,width=1.0\linewidth]{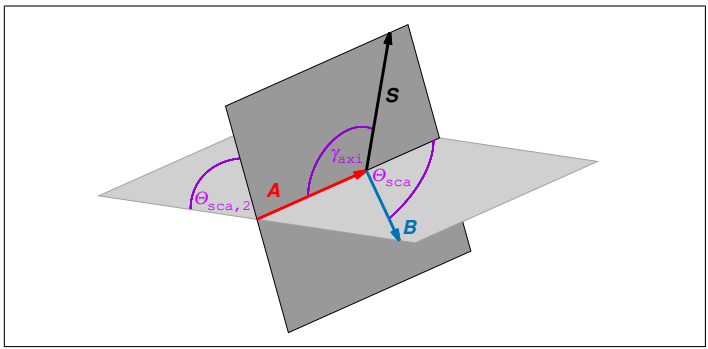}}
\caption{Sketch of a general scattering event on an oblate dust grain, including the incidence direction $\vec{A}$, the scattering direction $\vec{B}$ and the symmetry axis $\vec{S}$ of the oblate grain. The angles $\gamma_\text{axi}$, $\Theta_\text{sca}$ and $\Theta_\text{sca,2}$ determine the scattering plane and the plane defined by $\vec{A}$ and $\vec{S}$.}
\label{fig_sketch_elong} 
\end{figure}

In order to describe a scattering event on an elongated dust grain, three axes are important: the axis parallel to the direction of the incidence radiation, $\vec{A}$,  the axis parallel to the direction of the scattered radiation, $\vec{B}$, and the symmetry axis of the grain, $\vec{S}$ (Fig.~\ref{fig_sketch_elong}). The incidence direction $\vec{A}$ and the symmetry axis $\vec{S}$ span  the angle $\gamma_\text{axi}$. 
For $\gamma_\text{axi} = 90^\circ$, the symmetry axis $\vec{S}$ is perpendicular to the incidence direction $\vec{A}$. The scattering direction is  defined by the two scattering angles $\Theta_\text{sca}$ and $\Theta_\text{sca,2}$, where the former is the angle between the incidence direction $\vec{A}$ and the scattering direction $\vec{B}$, and  $\Theta_\text{sca,2}$ is the angle between the scattering plane (defined by $\vec{A}$ and $\vec{B}$) and the plane defined by $\vec{A}$ and $\vec{S}$.

 \begin{figure}
 \fbox{\includegraphics[trim=0cm -0.33cm 0cm -0.33cm,  clip=true,page=1,width=1.0\linewidth]{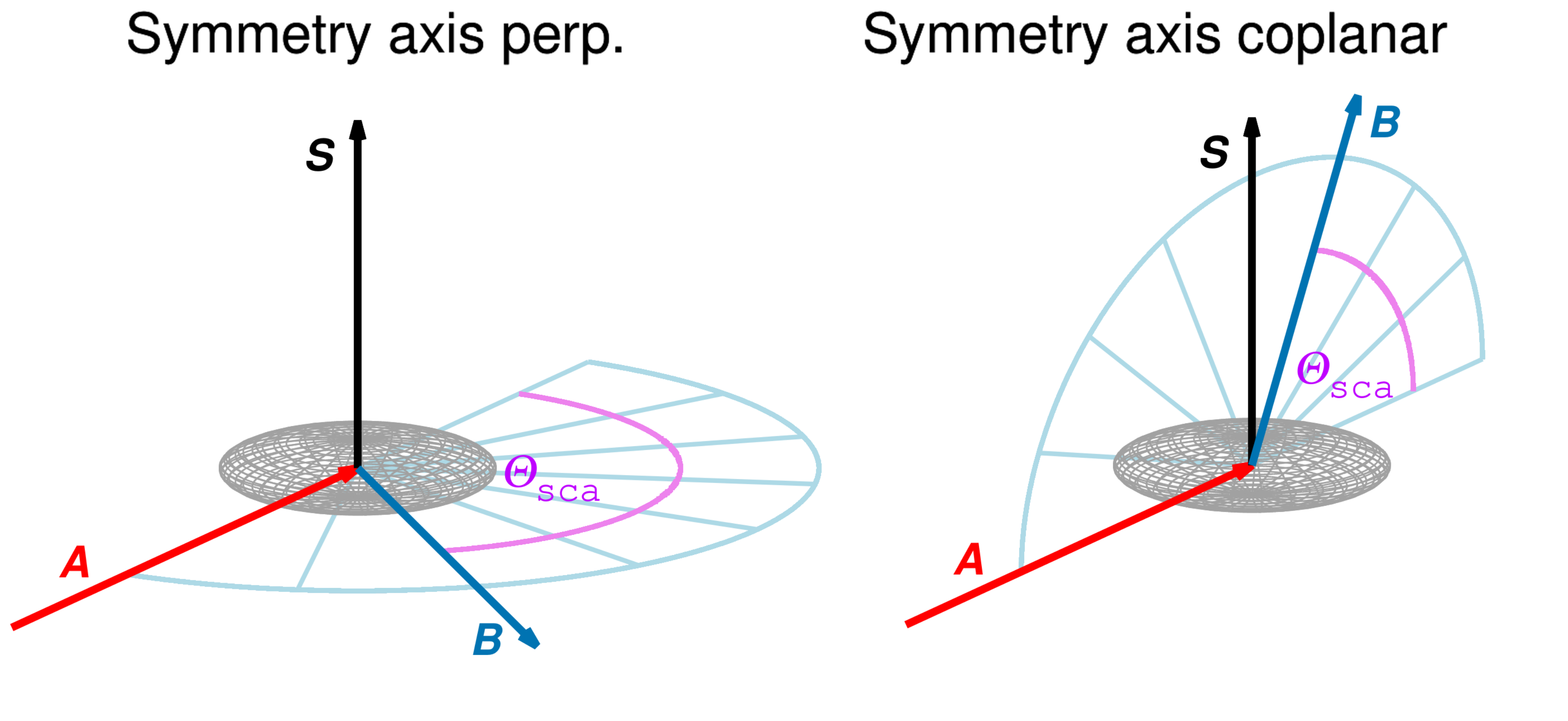}}
\caption{Orientation of the two considered scattering planes (blue grid). \textit{Left}: Grain symmetry axis $\vec{S}$ is perpendicular to scattering plane ($\Theta_\text{sca,2}=90^\circ$). \textit{Right}: Symmetry axis $\vec{S}$ is coplanar to scattering plane ($\Theta_\text{sca,2}=0^\circ$).}
\label{fig_scat_plane} 
\end{figure}

The polarisation degree of the incident and scattered radiation are $P_\text{ini}$ and $P_\text{sca}$, respectively, and the angle of polarisation of the incident radiation is $\gamma_\text{ini}$. For $\gamma_\text{ini}=0^\circ$, the incident polarisation orientation is parallel to the long semi-axis $c$ of the oblate grain (perpendicular to $\vec{S}$). The polarisation degree $P_\text{sca}$ of the scattered radiation is positive when the polarisation orientation is perpendicular to the scattering plane, and $P_\text{sca}$ is negative when it is coplanar to it. A transition from positive to negative polarisation is accompanied by a $\unit[90]{degree}$-flip of the polarisation orientation that is known as the polarisation reversal (e.g.,~\citealt{Daniel1980, Kirchschlager2014, Brunngraeber2019}).
 
We calculated the optical properties for different grain sizes $a_\text{eff}$, axis ratios $\nicefrac{c}{b}$, and scattering angles $\Theta_\text{sca}$ and $\Theta_\text{sca,2}$. The dust material is astronomical silicate (\citealt{Draine2003a,Draine2003b}) and the wavelength amounts to $\unit[870]{\mu m}$ which represents the ALMA waveband B7. The incidence angle is set to $\gamma_\text{axi}=90^\circ$ so that the incident radiation shines on the edge of the oblate grain and incoming photons ``see'' an ellipsoidally geometric cross section of the oblate grain with semi-axis $b$ and $c$.
We investigate the scattering along two orientations of the scattering plane, $\Theta_\text{sca,2}=90^\circ$ (grain symmetry axis $\vec{S}$ perpendicular to the scattering plane, Fig.~\ref{fig_scat_plane}, \textit{left}) and $\Theta_\text{sca,2}=0^\circ$ ($\vec{S}$ coplanar to scattering plane, Fig.~\ref{fig_scat_plane}, \textit{right})\footnote{For convenience, we limit our study to these two scattering planes which constitute special cases.}. 

In Sections~\ref{301} and ~\ref{401} the results are presented for dust grains with effective radii \mbox{$a_\text{eff}=\unit[100]{\mu m}$} and \mbox{$a_\text{eff}=\unit[150]{\mu m}$}. We chose these grain sizes as they lie within the Rayleigh regime (\mbox{$\unit[100]{\mu m}$}), i.e. the grain size is much smaller than the wavelength ($2\pi a_\text{eff}/\lambda < 1$), or already in the Mie~regime (\mbox{$\unit[150]{\mu m}$}) where the mean grain size approaches the
wavelength.


\section{Scattering on non-spherical dust grains}
\label{301}
We begin with a simplified setup in which an unpolarised incoming wave scatters at a non-spherical grain (for a polarised incoming wave, see Section~\ref{401}). We present this scenario for a single-size grain as well as a grain size distribution and compare our results to the case of self-scattering on spherical grains. 

  \begin{figure*}
     \centering
     \includegraphics[trim=1.2cm 0.3cm 2.4cm 1.1cm,  clip=true, width=0.5\linewidth, page=1]{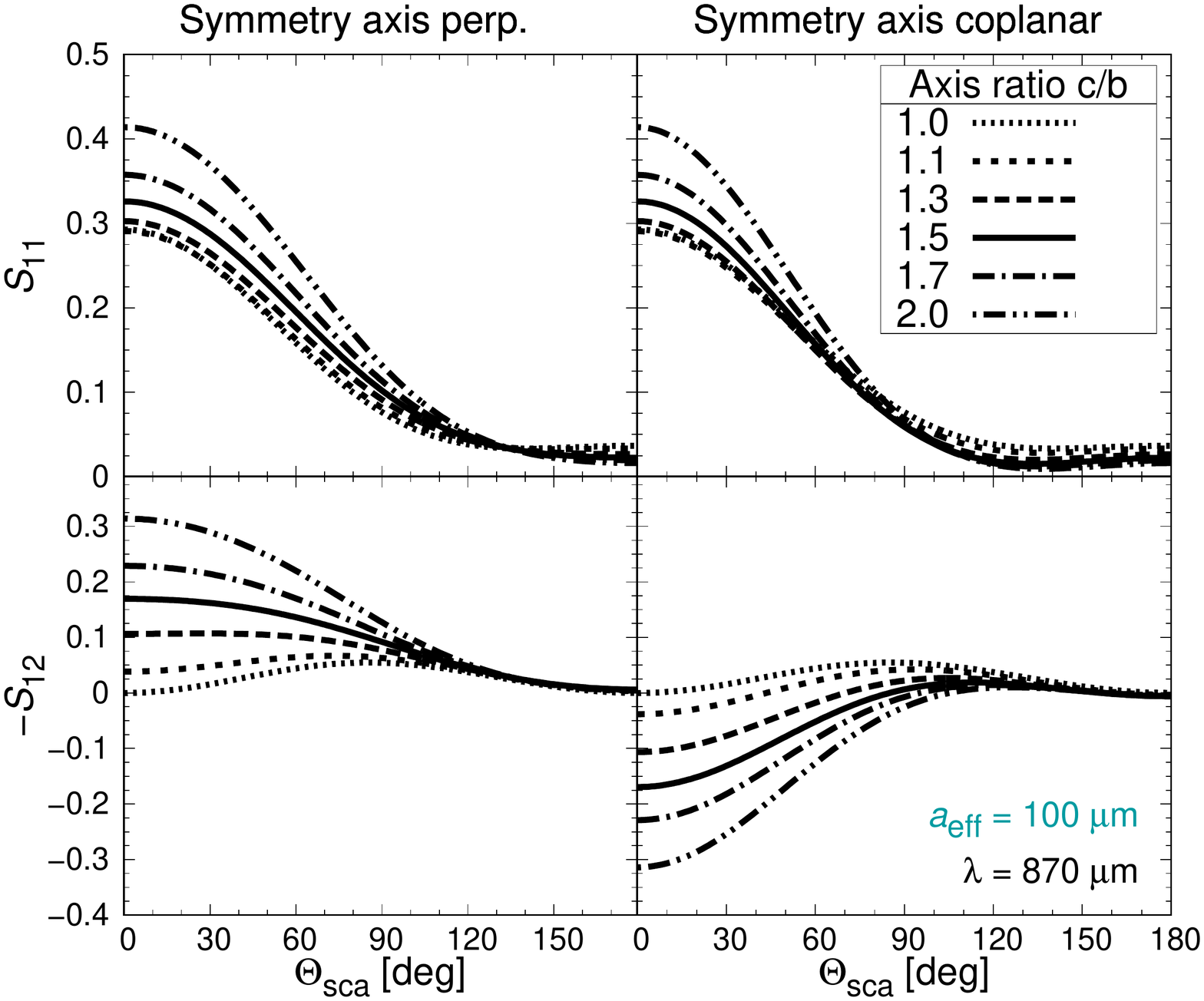}\hspace*{-0.1cm}
     \includegraphics[trim=1.2cm 0.3cm 2.4cm 1.1cm,  clip=true, width=0.5\linewidth, page=1]{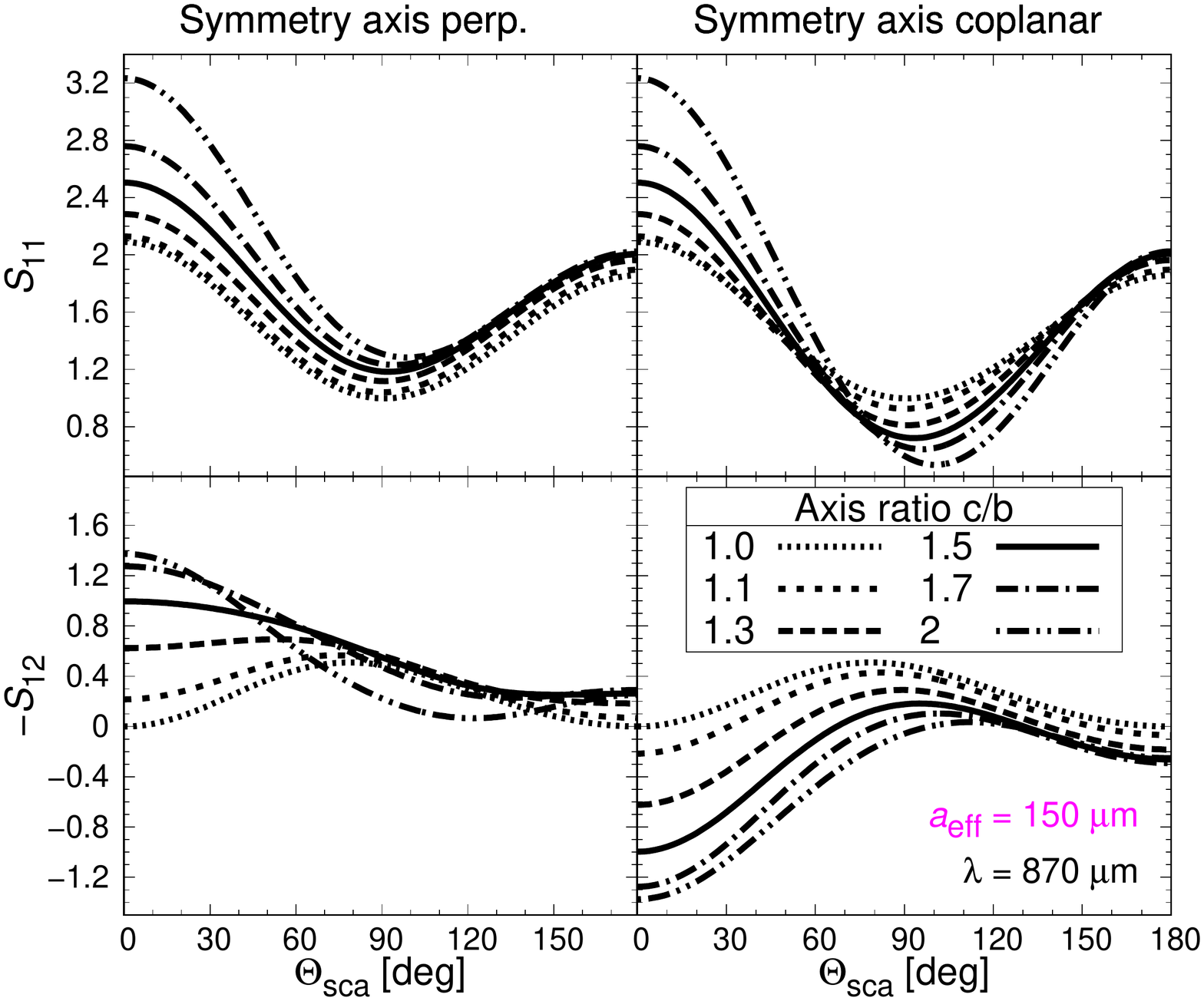}
        \caption{Scattering matrix elements $S_{11}$ and ($-S_{12}$) of oblate grains as a function of the scattering angle $\Theta_\text{sca}$ and different axis ratios $\nicefrac{c}{b}$, for the two scattering planes perpendicular and coplanar to the grain's symmetry axis ($\Theta_\text{sca,2}=90^\circ$ and $0^\circ$, respectively). The grain size is \mbox{$a_\text{eff}=\unit[100]{\mu m}$} (Rayleigh~regime; \textit{left}) and \mbox{$a_\text{eff}=\unit[150]{\mu m}$} (Mie~regime; \textit{right}).}
           \label{fig_sca}
     \end{figure*}

\subsection{Scattering elements of single-size grains}
\label{333}    
     
     \begin{figure*}
     \centering
          \includegraphics[trim=0.0cm 0.0cm 0.0cm 0.0cm,  clip=true, width=1.0\linewidth, page = 1]{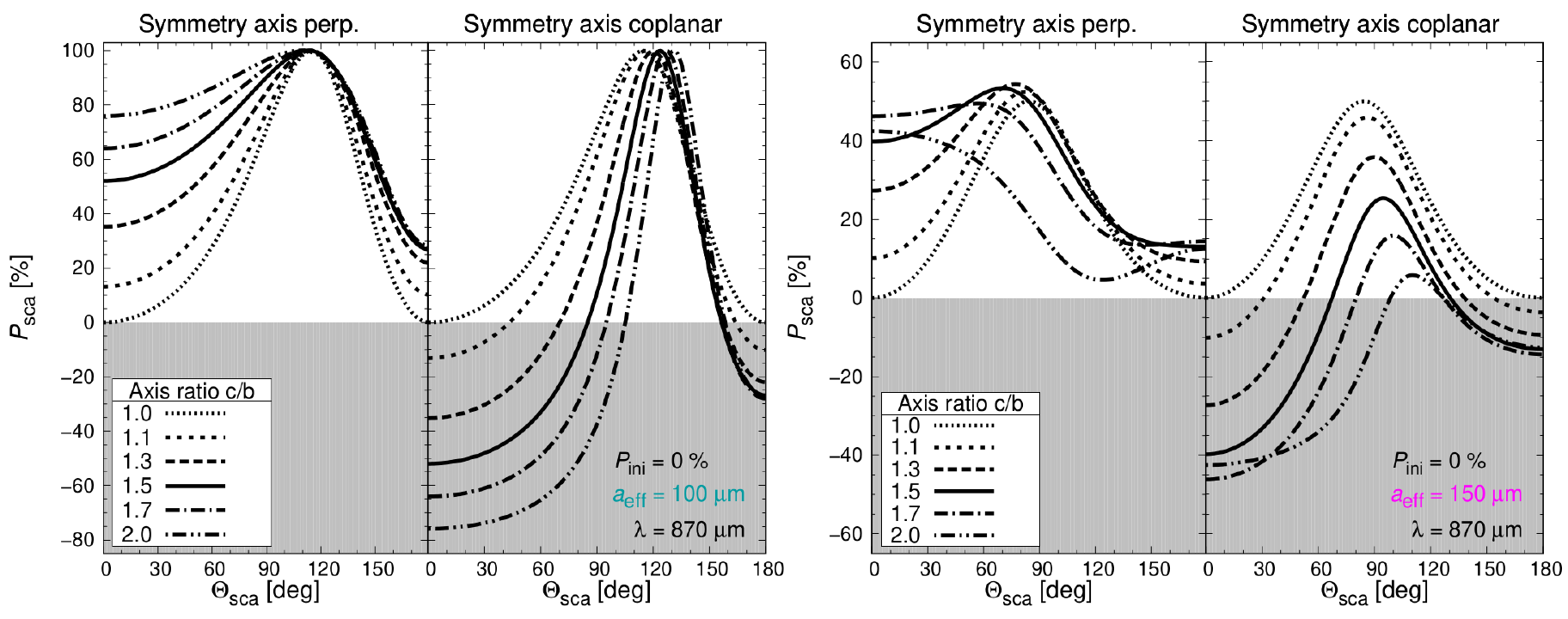}    
          \caption{Degree of polarisation $P_\text{sca}$ by scattering on oblate dust grains as a function of the scattering angle $\Theta_\text{sca}$ and different axis ratio $\nicefrac{c}{b}$, for the two scattering planes perpendicular and coplanar to the grain's symmetry axis ($\Theta_\text{sca,2}=90^\circ$ and $0^\circ$, respectively). The polarisation $P_\text{ini}$ of the incident radiation is zero and the grain size \mbox{$a_\text{eff}=\unit[100]{\mu m}$} (\textit{left}) and \mbox{$a_\text{eff}=\unit[150]{\mu m}$} (\textit{right}). The grey-shaded region indicates negative scattering polarisations which imply a reversal of the polarisation orientation.}                                    
           \label{fig_ips1}                           
     \end{figure*}             
     
The scattering matrix elements $S_{11}$ and $(-S_{12})$ of dust grains with effective radius \mbox{$a_\text{eff}=\unit[100]{\mu m}$} and \mbox{$a_\text{eff}=\unit[150]{\mu m}$} are shown in Fig.~\ref{fig_sca} as a function of the scattering angle $\Theta_\text{sca}$ and the axis ratio $\nicefrac{c}{b}$. 

For the case of scattering on the $\unit[100]{\mu m}$ grains (Fig.~\ref{fig_sca}, \textit{left}) perpendicular to the symmetry axis~$\vec{S}$ ($\Theta_\text{sca,2}=90^\circ$), $S_{11}$ and $(-S_{12})$ are increasing with increasing axis ratio $\nicefrac{c}{b}$ when the scattering angles are below ${\sim}120^\circ$. This indicates an enhanced forward scattering of oblate grains. For larger scattering angles, the impact of the axis ratio on the scattering matrix elements is small. For the case that the symmetry axis $\vec{S}$ is coplanar to the scattering plane ($\Theta_\text{sca,2}=0^\circ$), $(-S_{12})$ shows an opposite trend compared to the scattering perpendicular to $\vec{S}$ while $S_{11}$ is mostly unaffected. The angular averaged value of $S_{11}$ is for $\Theta_\text{sca,2}=90^\circ$ slightly larger than for $\Theta_\text{sca,2}=0^\circ$ which implies a higher scattering efficiency ($C_\text{sca}$) along the long axis compared to the short axis (cf. \citealt{Cho2007}). Compared to the \mbox{$\unit[100]{\mu m}$} grains, the values of $S_{11}$ of the \mbox{$\unit[150]{\mu m}$} grains (Fig.~\ref{fig_sca}, \textit{right}) increase by a factor of six and indicate a rising scattering behaviour at scattering angles larger than $90^\circ$ (backward scattering enhancement). The values of $(-S_{12})$ roughly increase by a factor of 4 and show a stronger dependence on the grain axis ratio for scattering angles larger than ${\sim}120^\circ$. 

The significant differences of the scattering behaviour of  \mbox{$\unit[150]{\mu m}$} grains compared to that of \mbox{$\unit[100]{\mu m}$} grains is representative for grains \mbox{$\unit[{>}100]{\mu m}$}, illustrating that they are beyond the Rayleigh regime, i.e., in the Mie regime at an observing wavelength of \mbox{$\lambda=\unit[870]{\mu m}$}.

\subsection{Linear polarisation of single-size grains}

The degree of linear polarisation by scattering, 
\begin{align}
P_\text{sca}=-S_\text{12}/S_\text{11}
 \label{eq01}
\end{align}
(e.g.,~\citealt{BohrenHuffman83}), reveals quantitative differences between oblate and spherical 
\mbox{$\unit[100]{\mu m}$} grains (Fig.~\ref{fig_ips1},~\textit{left}). The scattered wave perpendicular to the symmetry axis~$\vec{S}$ (\mbox{$\Theta_\text{sca,2}=90^\circ$}) possesses a polarisation that increases with increasing axis ratio $\nicefrac{c}{b}$, in particular for scattering angles below $100^\circ$ but also for large angles ($\Theta_\text{sca}\gtrsim150^\circ$). This implies that both forward and backward scattering have an enhanced polarisation fraction in the case of oblate dust grains. The peak polarisations amount to $\unit[100]{\%}$ for all axis ratios and occur at $\Theta_\text{sca}{\sim}110^\circ$. For the case that the symmetry axis $\vec{S}$ is coplanar to the scattering plane ($\Theta_\text{sca,2}=0^\circ$), the scattering polarisation is decreasing with increasing axis ratio. Negative scattering polarisations at small ($\Theta_\text{sca}\lesssim 60^\circ-90^\circ$) and large scattering angles ($\Theta_\text{sca}\gtrsim150^\circ$) occur in combination with a 90 degree flip of the polarisation orientation. The peak polarisation is also $\unit[100]{\%}$  for all axis ratios but shifts from $\Theta_\text{sca}{\sim}110^\circ$ for spherical grains to ${\sim}130^\circ$ for oblate grains ($\nicefrac{c}{b}=2.0$).

The scattering polarisations at $\Theta_\text{sca}=0^\circ$ and $\Theta_\text{sca}=180^\circ$ of both scattering planes are for reasons of symmetry equal to each other. The different polarisation signs in Fig.~\ref{fig_ips1} (and also in Figs.~\ref{fig_ips2} and \ref{fig_ips3}) at these scattering angles result from the $90^\circ$ rotation of the reference system.

\cite{Kataoka2015} shows for grain size distributions of spherical grains, composed of a mixture of silicate, water ice, and organics, that the maximum polarisation strongly drops from $\unit[100]{\%}$ to $\unit[{\ll}10]{\%}$ when the maximum grain size is increased from $\unit[100]{\mu m}$ to $\unit[300]{\mu m}$ (for a wavelength of $\unit[870]{\mu m}$). The significant changes in scattering polarisation when increasing the grain size can also be seen for non-spherical silicates (Fig.~\ref{fig_ips1}, \textit{right}, for $a_\text{eff}=\unit[150]{\mu m}$). These grains are already in the Mie regime. The peak polarisation as well as the corresponding scattering angle $\Theta_\text{sca}$ at which the maximum polarisation occurs depend on the grain axis ratio $\nicefrac{c}{b}$.
For scattering perpendicular to the symmetry axis $\vec{S}$ ($\Theta_\text{sca,2}=90^\circ$), the peak polarisation reaches only values up to $\unit[55]{\%}$, increases with axis ratio if $\nicefrac{c}{b}\le1.3$ and decreases for larger axis ratios. The  scattering angle of the peak polarisation is shifted from $\Theta_\text{sca}{\sim}90^\circ$ for spherical grains to $0^\circ$ for oblate grains ($\nicefrac{c}{b}=2.0$). For the case that $\vec{S}$ is coplanar to the scattering plane ($\Theta_\text{sca,2}=0^\circ$), the scattering polarisation is monotonic decreasing with increasing axis ratio. As a consequence, the scattering polarisation of oblate grains with $\nicefrac{c}{b}=2.0$ is negative for nearly all scattering angles except around the peak polarisation. The  scattering angle of the peak polarisation shifts from $\Theta_\text{sca}{\sim}90^\circ$ for spherical grains to $110^\circ$ for oblate grains ($\nicefrac{c}{b}=2.0$). 

In conclusion, spherical grains in the Rayleigh regime ($2\pi a_\text{eff}/\lambda < 1$) show quantitative differences compared to oblate grains, but they are still a convenient representative. However, for self-scattering on larger grains (Mie regime) the spherical dust model shows significant qualitative deviations and has to be clearly distinguished to that of non-spherical grains. Moreover, when the wavelength is no longer fixed but increased, the Rayleigh regime is shifted to larger and larger grain sizes (see e.g. Appendix \ref{waves}).

\subsection{Circular polarisation of single-size grains}
\label{circ}
Finally, we also studied the amount of circular polarisation, 
\begin{align}
P_\text{sca,circ}=S_\text{41}/S_\text{11},
 \label{eq02}
\end{align}
of scattered, initially unpolarised radiation on oblate grains.
Similar to spherical grains, \mbox{$P_\text{sca,circ}$} is negligible for all considered scattering angles, axis ratios and grain radii as the scattering matrix element $|S_\text{41}|$ is at least six orders of magnitude below $|S_\text{11}|$. 
The reason is the symmetric orientation of the symmetry axis ($\gamma_\text{axi}=90^\circ$) and the scattering planes ($\Theta_\text{sca,2}=0^\circ$ and $90^\circ$). Non-orthogonal or non-parallel orientations of the axes cause significant circular polarisations even for initially unpolarised radiation (e.g.,~\citealt{Gledhill2000}). Moreover, a significant circular polarisation is also expected if the incoming wave is intrinsically polarised (see Section \ref{43}).


\subsection{Linear polarisation of a grain size distribution}
The linear polarisation $P_\text{sca}$ by scattering on single-size spherical grains shows a strong wavelength dependence for \mbox{$2\pi a_\text{eff}/\lambda$>1} with a high frequency and strong oscillations (see, e.g., Fig.~2 in \citealt{Brunngraeber2019}). The oscillations enable negative scattering polarisations with a reversal of the polarisation orientation. 
For a fixed wavelength, the polarisation oscillations occur also as a function of grain size. We expect that the oscillations could cancel out for an ensemble of different grain sizes or a continuous distribution of grain sizes as the contribution of positive and negative (or the contribution of parallel and perpendicular) scattering polarisations annihilates each other (e.g.,~\citealt{Kirchschlager2019}).\footnote{We note that the mutual obliteration of polarised signals of grains of different radii is the reason for the relatively smooth polarisation pattern in Fig.~3 in \cite{Kataoka2015} where the polarisation is displayed as a function of maximum radius of a size distribution of spherical grains.} 

   \begin{figure}
 \centering
     \includegraphics[trim=2.5cm 2.6cm 2.1cm 2.5cm,  clip=true, width=1.0\linewidth, page = 1]{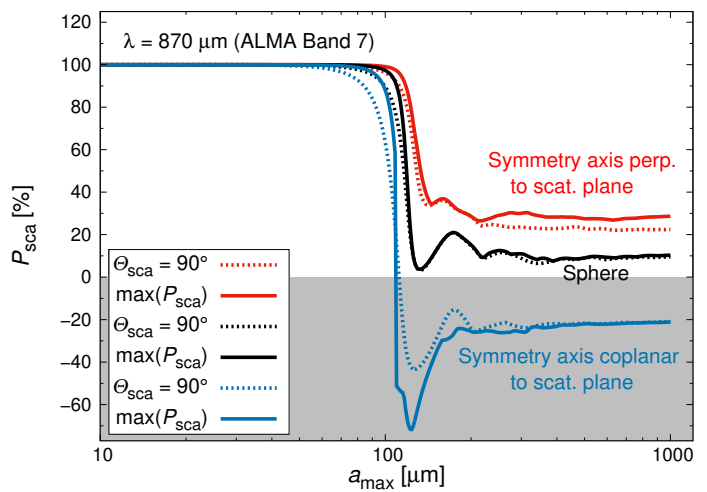}
 \caption{Scattering polarisation $P_\text{sca}$ for a size distribution of oblate (red and blue) and spherical dust grains (black) as a function of the maximum grain size $a_\text{max}$. Shown is the peak polarisation of the scattering interval $\left[0^\circ,180^\circ\right]$ (solid lines) as well as the polarisation at $\Theta_\text{sca}=90^\circ$ (dotted lines). The size distribution follows $\text{d}n\propto a_\text{eff}^{-3.5}\text{d}a_\text{eff}$ and the axis ratio is $\nicefrac{c}{b}=1.5$.}                                    
 \label{fig_distrib}                           
 \end{figure}

 \begin{figure*}                                   
\centering                                       
\includegraphics[trim=0.0cm 0cm 0.0cm 0.0cm,  clip=true, width=1.0\linewidth, page = 1]{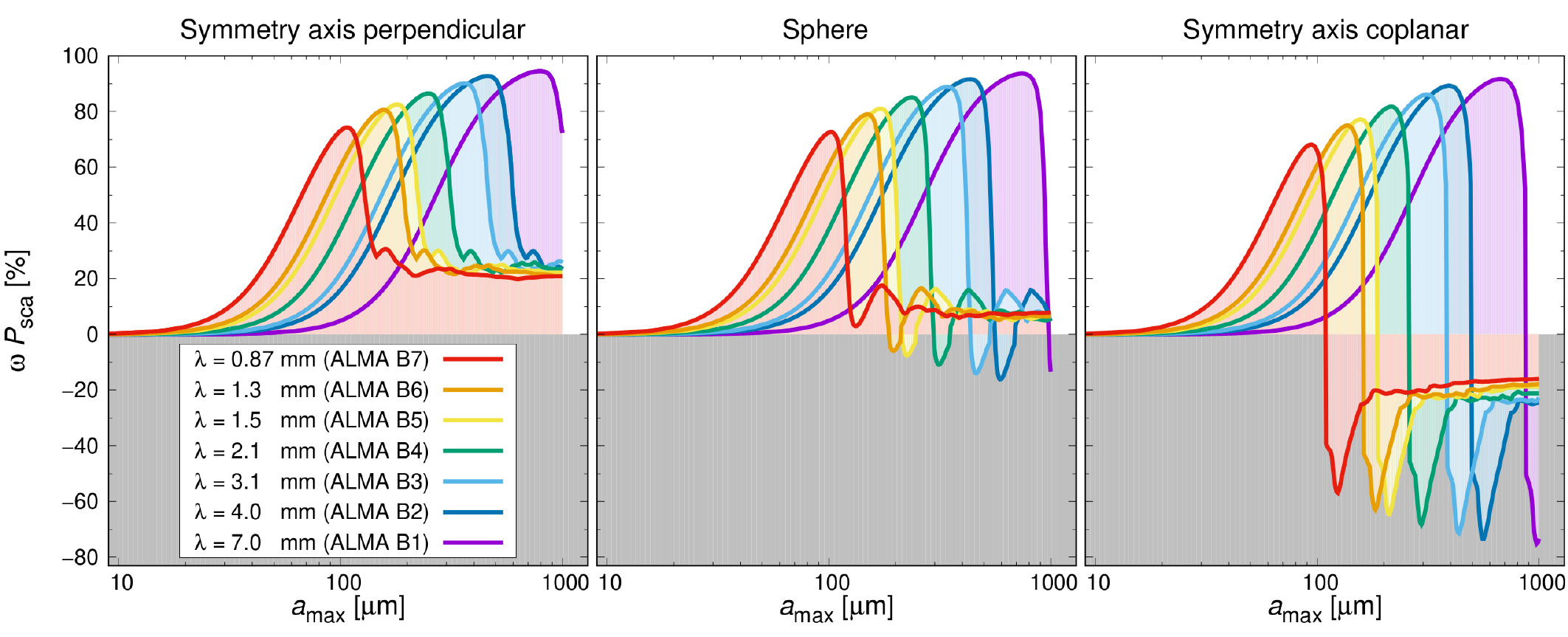}
\caption{Albedo $\omega$ times scattering polarisation $P_\text{sca}$ for a grain size distribution $\text{d}n\propto a_\text{eff}^{-3.5}\text{d}a_\text{eff}$ as a function of the maximum grain size $a_\text{max}$. 
This figure represents the grain size that contributes most to the polarised intensity. 
\textit{Left}: Scattering on oblate grains ($\nicefrac{c}{b}=1.5$) perpendicular to the grain's symmetry axis ($\Theta_\text{sca,2}=90^\circ$). 
\textit{Center}: Scattering on spherical grains ($\nicefrac{c}{b}=1.0$) . 
\textit{Right}: Scattering on oblate grains ($\nicefrac{c}{b}=1.5$), the grain's symmetry axis is coplanar to the scattering plane ($\Theta_\text{sca,2}=0^\circ$).}                                      
\label{fig_window}                             
\end{figure*}  

In order to study the scattering polarisation of grain size distributions of oblate dust grains, we calculated the optical properties for 1000 grain sizes $a_\text{eff}$ which are logarithmically equidistantly distributed in the interval $\left[\unit[5]{nm},\unit[1]{mm}\right]$. Grain axis ratio and wavelength are fixed to $\nicefrac{c}{b}=1.5$ and $\lambda=\unit[870]{\mu m}$, respectively. Subsequently, the scattering polarisation is calculated for a size distribution $\text{d}n\propto a_\text{eff}^{-3.5}\text{d}a_\text{eff}$ with minimum radius $a_\text{min}=\unit[5]{nm}$ while the maximum grain size is varied between $\unit[10]{\mu m}$ and $\unit[1]{mm}$, and the peak polarisation (maximum polarisation as a function of scattering angle $\Theta_\text{sca}$) is determined (Fig.~\ref{fig_distrib}). We assume that all grains are perfectly aligned as otherwise the contributions of differently aligned grains would mitigate each other or even cancel each other out.

We distinguish between the scattering plane perpendicular and coplanar to the symmetry axis $\vec{S}$ (see Fig.~\ref{fig_scat_plane}). For both, the peak polarisation is $\unit[100]{\%}$ and the scattering angle of the peak polarisation $\Theta_\text{sca}{\sim}90^\circ$ when the maximum grain size $a_\text{max}$ is small compared to the wavelength (Rayleigh regime). The polarisation changes significantly for grain sizes slightly larger than $a_\text{max}=\unit[100]{\mu m}$ as they are already in the Mie regime: the peak polarisation of scattering perpendicular to~$\vec{S}$ ($\Theta_\text{sca,2}=90^\circ$) drops down to $\unit[{\sim}35]{\%}$ for $a_\text{max}=\unit[150]{\mu m}$, shows some moderate variations for larger sizes before converging against a constant polarisation degree of $\unit[{\sim}29]{\%}$ at grain sizes up to $\unit[1]{mm}$. The peak polarisation of the scattering plane coplanar to $\vec{S}$ ($\Theta_\text{sca,2}=0^\circ$) even drops to $\unit[{\sim}-70]{\%}$ for $a_\text{max}=\unit[120]{\mu m}$ and converges to $\unit[{\sim}-21]{\%}$ for larger grain sizes, and the peak polarisation is flipped in orientation by $90^\circ$ for all size distributions with $a_\text{max}\gtrsim\unit[110]{\mu m}$.

We also calculated the scattering polarisation for a size distribution of spherical dust grains using DDA (Fig.~\ref{fig_distrib}). The polarisation degrees of the spherical grains are larger than that of scattering along the short axis (coplanar scattering) and lower than that of scattering along the long axis (perpendicular to symmetry axis $\vec{S}$). The scattering polarisation of the spherical grains is $\unit[100]{\%}$ for small grains and starts decreasing to $\sim\unit[0-10]{\%}$ when $a_\text{max}\gtrsim\unit[100]{\mu m}$, which is comparable to the result of \cite{Kataoka2015}. We note that \cite{Kataoka2015} used a different dust composition made of silicate, water ice, and organics.

Independent of the grain axis ratio or scattering plane, the scattering angle of the peak polarisation shows strong oscillations and discontinuous jumps due to the occurrence of multiple polarisation maxima for grain sizes larger than $\unit[{\sim}100]{\mu m}$. The scattering angles of the peak polarisation strongly deviate from $\Theta_\text{sca}=90^\circ$ and cover the full  interval from $0^\circ$ to $180^\circ$. We plotted in Fig.~\ref{fig_distrib} also the scattering polarisation at $\Theta_\text{sca}=90^\circ$ which shows absolute differences of up to $\unit[30]{\%}$ compared to the peak polarisation values.    

Furthermore, we calculated the scattering polarisation of grain size distributions for the different ALMA wavebands B6, B5, B4, B3, B2, and B1 ($\lambda=\unit[1.3]{mm},\unit[1.5]{mm},\unit[2.1]{mm}, \unit[3.1]{mm},\unit[4]{mm}$ and $\unit[7]{mm}$; see Appendix \ref{waves}). The Rayleigh limit is linearly shifted with wavelength to larger maximum grain sizes while the differences between the polarisation degree and orientation of spherical and non-spherical grains are preserved. 

Fig.~\ref{fig_window} shows the product of the scattering polarisation $P_\text{sca}$ and the albedo $\omega=\nicefrac{\kappa_\text{sca}}{(\kappa_\text{abs}+\kappa_\text{sca})}$ where $\kappa_\text{abs}$ and $\kappa_\text{sca}$ are the absorption and scattering mass opacity of the grain size distributions, respectively. Contrary to the scattering polarisation, the albedo is almost zero for $a_\text{max}\ll  \lambda/2\pi$ and increases with increasing grain size. Following the approach of \cite{Kataoka2015} for spherical grains, the product of albedo and scattering polarisation,  $\omega\,P_\text{sca}$, defines a window function for the grain sizes that contribute to the scattering polarisation: only grain size distributions with maximum radii $a_\text{max}\sim \lambda/ 2\pi$ have both a significant scattering polarisation and albedo and dominate the polarisation signal at \mbox{(sub-)}millimetre wavelengths.  The window functions of non-spherical (Fig.~\ref{fig_window}, \textit{left} and \textit{right}) and spherical (Fig.~\ref{fig_window}, \textit{center}) dust grains are very similar to each other for maximum sizes smaller than or comparable to $\lambda/2\,\pi$, indicating that grains of size of at least ${\sim}\lambda/ 2\pi$ are required to account for polarisation levels of a few per cent detected  in several (sub-)millimetre observations. Furthermore, the window functions of non-spherical and spherical dust grains peak at the same grain radii and the peak polarisations increase with increasing wavelength.

 \begin{figure*}                                   
\centering                                       
\includegraphics[trim=0.6cm 1.5cm 2.4cm 1.05cm,  clip=true, width=0.5\linewidth, page = 2]{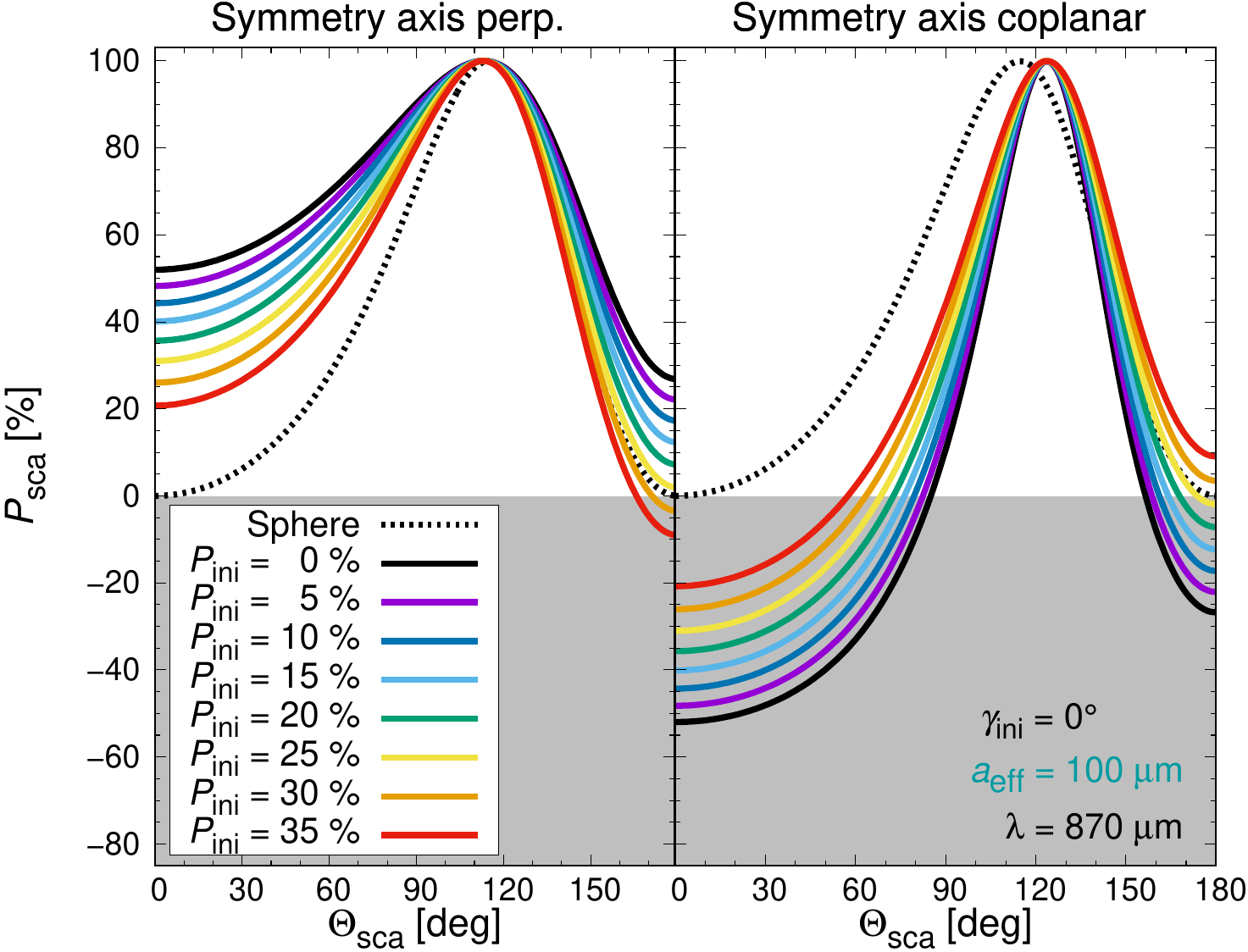}\hspace*{-0.1cm}
\includegraphics[trim=0.6cm 1.5cm 2.4cm 1.05cm,  clip=true, width=0.5\linewidth, page = 2]{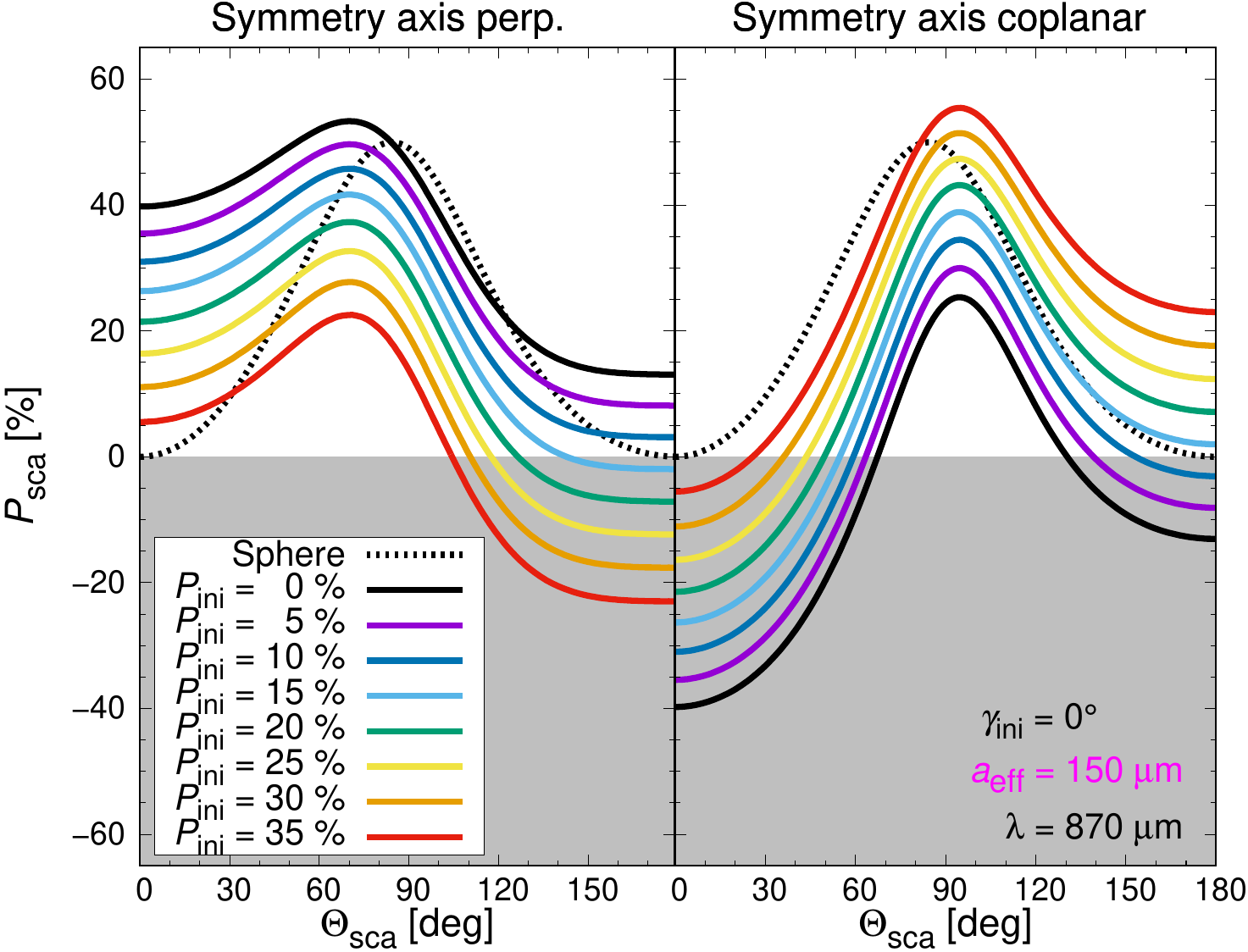}
\caption{Degree of polarisation $P_\text{sca}$ by scattering on non-spherical (oblate) dust grains as a function of the scattering angle $\Theta_\text{sca}$ and the polarisation $P_\text{ini}$ of the incident radiation. The incident polarisation orientation
is fixed to \mbox{$\gamma_\text{ini}=0^\circ$} (perpendicular to the symmetry axis $\vec{S}$) and the grain axis ratio is $\nicefrac{c}{b}=1.5$.  The grain size is \mbox{$a_\text{eff}=\unit[100]{\mu m}$} (\textit{left}) and \mbox{$a_\text{eff}=\unit[150]{\mu m}$} (\textit{right}). The scattering polarisation of an initially unpolarised wave on spherical grains is shown for comparison (black dotted lines).}                                      
\label{fig_ips2}                             
\end{figure*} 

The crucial differences between the window functions of non-spherical and spherical dust grains appear when the grain size approaches the wavelength. The amount of scattering polarisation of spherical grains drops down to a much lower value than for non-spherical grains (Fig.~\ref{fig_distrib}). A significant polarisation persists for large oblate grains which contribute to the self-scattering signal. Consequently, the grain sizes inferred from self-scattering observations are increased to values significantly larger than ${\sim}\lambda/ 2\pi$ when considering non-spherical grains. The self-scattering of non-spherical grains helps to reconcile the existing discrepancy between grain sizes  inferred from self-scattering and more indirect measurements as well as those expected from theory (e.g., \citealt{Beckwith1990}). The potential detection of a polarisation reversal in the self-scattering signals gives further information about whether the scattering is predominantly along the short or long axis of the spheroidal grains. 

Higher polarisation degrees can be realised by, amongst others, larger grain axis ratios $\nicefrac{c}{b}$. Thus, self-scattering on non-spherical dust grains allows for higher polarisation fractions than usually known from scattering on spherical grains. This is an important result as high polarisation values observed in disks have usually been interpreted as being produced by the polarised emission of elongated grains and not by self-scattering.

The presented results assume perfect grain alignment (all grain axes point into the same direction). The mean of the scattering polarisation of the two scattering planes ($\Theta_\text{sca,2}=90^\circ$ and $0^\circ$) of non-spherical grains is comparable to the scattering polarisation of spherical grains. Though an appropriate calculation of the scattering polarisation of randomly orientated dust grains needs more than two scattering planes, this already indicates that the oblate grains need to be well or nearly perfectly aligned in order to reveal the different self-scattering behaviour. Realistic grain alignment processes depend strongly on the surrounding gas and radiation conditions as well as on magnetic field strengths, which is well beyond the scope of this study and has to be postponed to future investigations.


\section{Self-scattering of intrinsically polarised waves}
\label{401}

 \begin{figure*}                                     
\centering                                         
\includegraphics[trim=0.6cm 1.5cm 2.4cm 1.05cm,  clip=true, width=0.5\linewidth, page = 3]{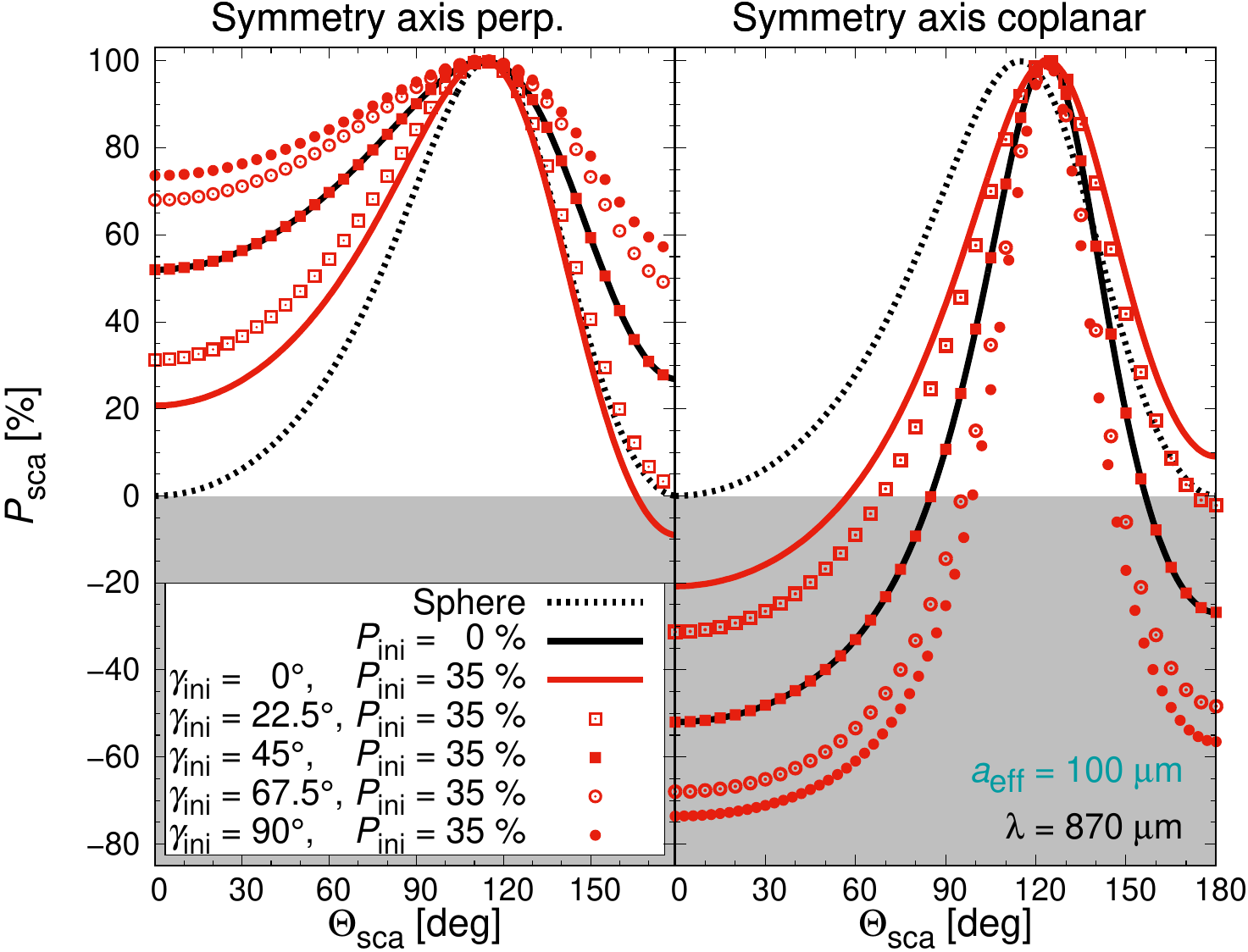}\hspace*{-0.1cm}
\includegraphics[trim=0.6cm 1.5cm 2.4cm 1.05cm,  clip=true, width=0.5\linewidth, page = 3]{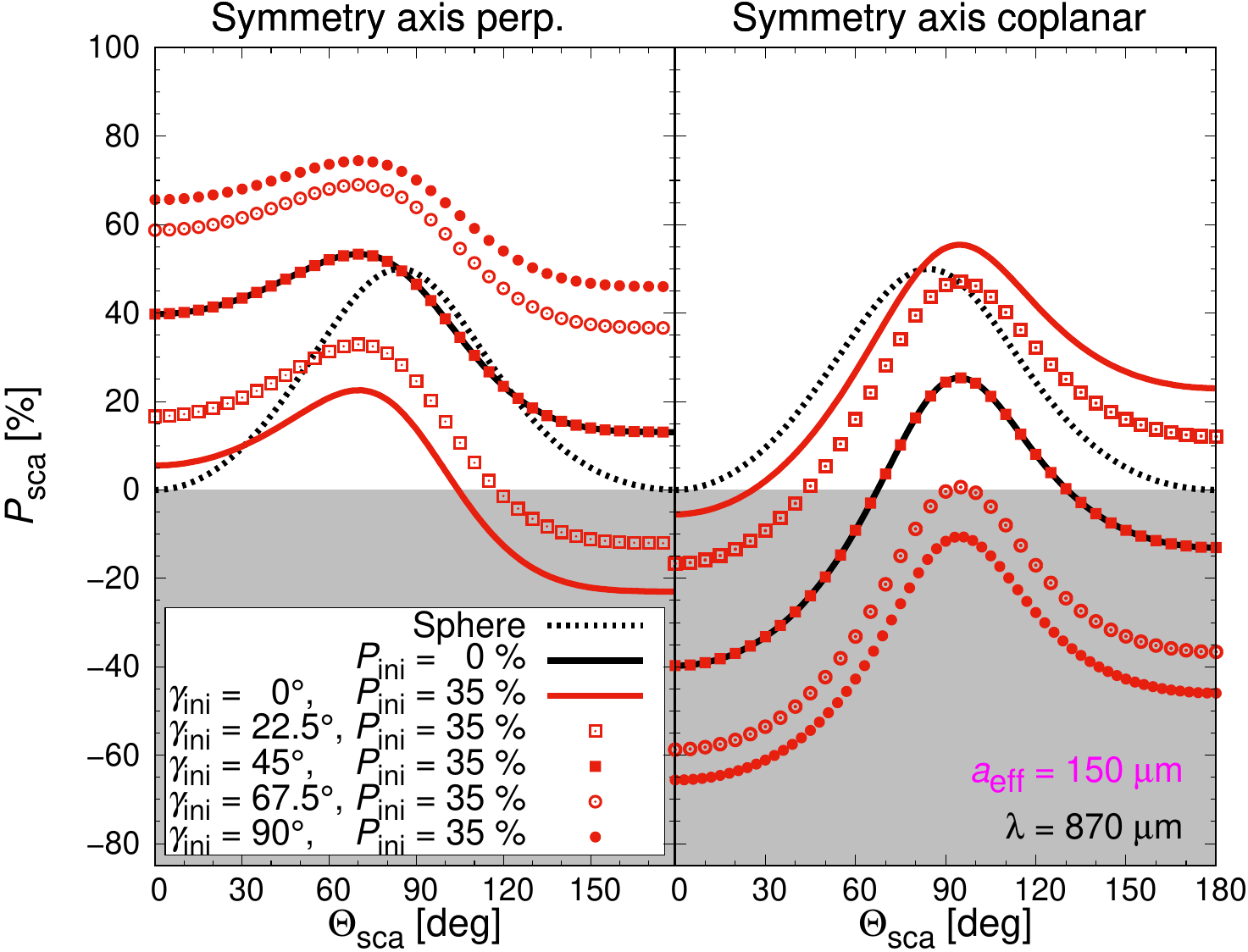}
\caption{Same as in Fig.~\ref{fig_ips2}, but for a fixed incident polarisation degree $P_\text{ini}=\unit[35]{\%}$ and different polarisation orientations $\gamma_\text{ini}$ of the incident radiation. The scattering polarisation of an initially unpolarised wave on spherical grains as well as on oblate grains is shown for comparison (black solid and dotted lines). \textit{Left}: \mbox{Grain size $a_\text{eff}=\unit[100]{\mu m}$}. \textit{Right:} \mbox{$a_\text{eff}=\unit[150]{\mu m}$}.
}
\label{fig_ips3}
\end{figure*}

In this section the self-scattering of intrinsically polarised radiation is investigated. As the optical properties of the grains at which the incoming wave gets scattered are unaffected by the polarisation state of the incident radiation, the scattering matrix elements from Section~\ref{301} can be reused and intrinsic polarisation is realised by adjusting the Stokes vector of the incoming wave (see Appendix~\ref{Stokes}). As in Section~\ref{301} the scattering polarisations for 
\mbox{$a_\text{eff}=\unit[100]{\mu m}$} and \mbox{$a_\text{eff}=\unit[150]{\mu m}$} grains are investigated as representatives for grains within the Rayleigh or Mie regime. To study the impact on the scattering polarisation, we varied the incident polarisation degree in Section~\ref{41} and the incident polarisation orientation in Section~\ref{42}. Circular polarisation is discussed in Section~\ref{43}.

\subsection{Scattering as a function of the incident polarisation degree $P_\text{ini}$}
\label{41}
We fix the grain axis ratio to $\nicefrac{c}{b}=1.5$ and the polarisation angle of the incident radiation to $\gamma_\text{ini}=0^\circ$ so that the incident polarisation is orientated perpendicular to the symmetry axis $\vec{S}$ (parallel to the long axis of the oblate grains).

The emission of non-spherical dust grains is intrinsically polarised (e.g., \citealt{Cho2007}). In \cite{Kirchschlager2019}, we show that a grain axis ratio of $\nicefrac{c}{b}=1.5$ results in a maximum intrinsic polarisation degree of $\unit[{\sim}35]{\%}$. The polarisation by emission is a potential source for the polarisation $P_\text{ini}$ of the incident radiation in an upcoming scattering event. In the following, we gradually increase the polarisation degree of the incident radiation, $P_\text{ini}$, from 0 to $\unit[35]{\%}$ and calculate the emerging scattering polarisation $P_\text{sca}$ (Fig.~\ref{fig_ips2}).

For initially unpolarised radiation, the orientation of the scattering polarisation is perpendicular to the scattering plane. Thus, the scattering polarisation is directed parallel to the symmetry axis $\vec{S}$ if the scattering is in the plane perpendicular to $\vec{S}$ ($\Theta_\text{sca,2}=90^\circ$). Increasing the incident polarisation $P_\text{ini}$ of the incident wave evokes a scattering polarisation component perpendicular to $\vec{S}$. Orthogonal contributions cancel each other out and the scattering polarisation parallel to $\vec{S}$ decreases with increasing $P_\text{ini}$. Consequently, intrinsically polarised radiation reduces the scattering polarisation when $\vec{S}$ is perpendicular to the scattering plane. 
For \mbox{$\unit[100]{\mu m}$} grains (Fig.~\ref{fig_ips2}, \textit{left}) and backward scattering angles ($\Theta_\text{sca}\gtrsim 150^\circ$), the orientation of the scattering polarisation can change by $90^\circ$ if the incident polarisation is higher than $P_\text{ini}\ge \unit[30]{\%}$. For \mbox{$\unit[150]{\mu m}$} grains (Fig.~\ref{fig_ips2}, \textit{right}), the polarisation flip occurs already for angles $\Theta_\text{sca}\gtrsim 100^\circ$ and incident polarisations $P_\text{ini}\ge \unit[15]{\%}$.

The trends are  inversed when the symmetry axis $\vec{S}$ is coplanar to the scattering plane ($\Theta_\text{sca,2}=0^\circ$) where the orientations of both the incident and the scattered polarisation are perpendicular to~$\vec{S}$.
For \mbox{$\unit[100]{\mu m}$} grains (Fig.~\ref{fig_ips2}, \textit{left}) the absolute amount of scattering polarisation is reduced for scattering angles \mbox{$\Theta_\text{sca}\lesssim 60^\circ-90^\circ$} and \mbox{$\Theta_\text{sca}\gtrsim 150^\circ$} (which means depolarisation), while it is increased for scattering angles around the peak polarisation \mbox{($60^\circ-90^\circ\lesssim\Theta_\text{sca}\lesssim 150^\circ$).} For \mbox{$\unit[150]{\mu m}$} grains (Fig.~\ref{fig_ips2}, \textit{right}), the absolute amount of scattering polarisation is reduced for scattering angles \mbox{$\Theta_\text{sca}\lesssim 30^\circ-70^\circ$} and \mbox{$\Theta_\text{sca}\gtrsim 120^\circ$,} while it is increased for scattering angles \mbox{$30^\circ-70^\circ\lesssim\Theta_\text{sca}\lesssim 120^\circ$.}

The scattering polarisation of \mbox{$a_\text{eff}=\unit[100]{\mu m}$} and \mbox{$a_\text{eff}=\unit[150]{\mu m}$} grains show further significant differences. The scattering polarisation phase functions of the \mbox{$\unit[150]{\mu m}$} grains are more asymmetric and the peak polarisations are smaller and shifted to smaller scattering angles ($\Theta_\text{sca}\sim70-90^\circ$). Moreover, the degree of the peak polarisation is decreasing with increasing intrinsic polarisation $P_\text{ini}$ while it is constant ($P_\text{sca}=\unit[100]{\%}$) for \mbox{$\unit[100]{\mu m}$} grains.

\begin{figure*}
\centering
 \includegraphics[trim=0.0cm 0.0cm 0.0cm 0.0cm,  clip=true, width=1.0\linewidth, page = 1]{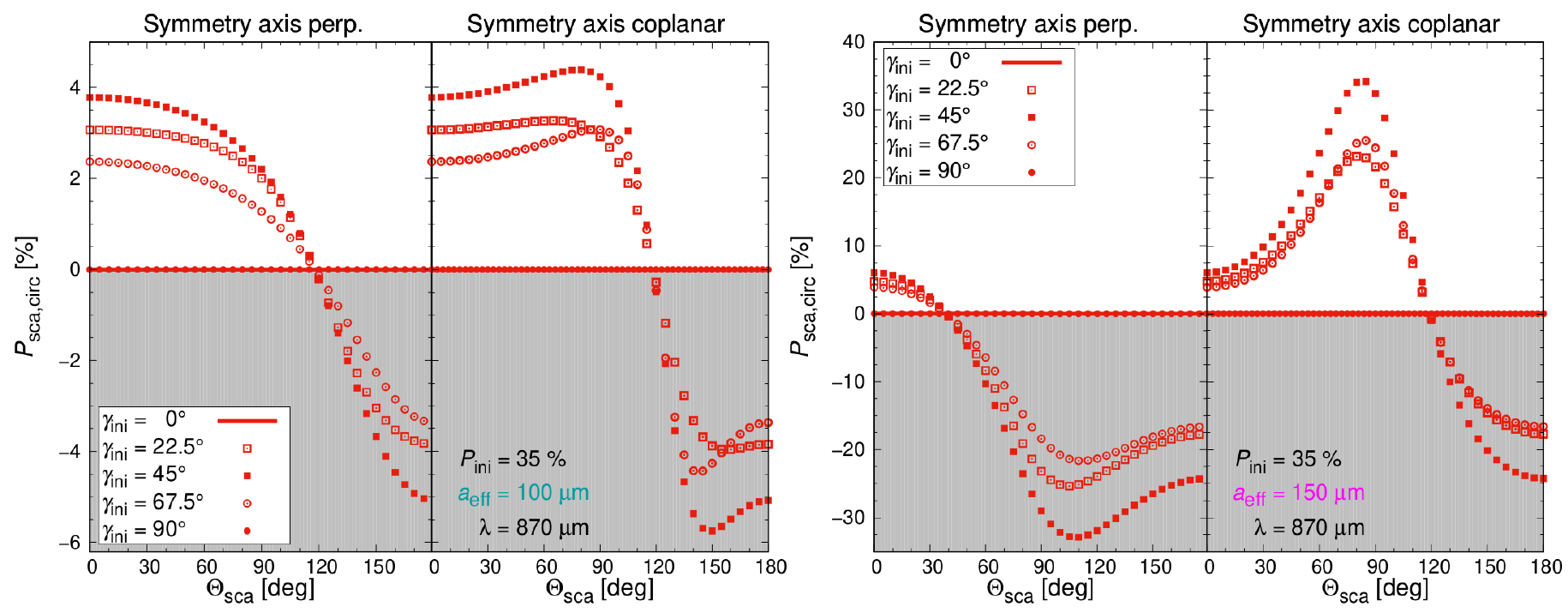}
\caption{Same as Fig.~\ref{fig_ips3}, only for circular polarisation $P_\text{sca,circ}$. \textit{Left}: \mbox{Grain size $a_\text{eff}=\unit[100]{\mu m}$}. \textit{Right:} \mbox{$a_\text{eff}=\unit[150]{\mu m}$}.
}                                    
\label{fig_circ}                           
\end{figure*}

\subsection{Scattering as a function of the orientation $\gamma_\text{ini}$ of incident polarisation}
\label{42}
In general, the polarisation angle of the incident radiation could have a random orientation with respect to the orientation of the non-spherical scattering grain, and we investigate in this section the influence of several different polarisation directions of the incident radiation. We gradually increased the initial polarisation angle from $\gamma_\text{ini}=0^\circ$ to $90^\circ$ with a step width of $22.5^\circ$ while the incident polarisation degree is fixed to $P_\text{ini}=\unit[35]{\%}$, and calculated the scattering polarisation $P_\text{sca}$ (Fig.~\ref{fig_ips3}).

Changing the polarisation angle $\gamma_\text{ini}$ of the incident radiation has the opposite effect to the increase of the incident polarisation degree (Section~\ref{41}). With increasing polarisation angle $\gamma_\text{ini}$, the proportion of incident polarisation that is perpendicular to the symmetry axis $\vec{S}$ mitigates and the proportion parallel to $\vec{S}$ rises. Consequently, an increase of the incident polarisation angle up to $90^\circ$ enhances the scattering polarisation in the case of $\Theta_\text{sca,2}=90^\circ$ and reduces it for $\Theta_\text{sca,2}=0^\circ$.  For $\gamma_\text{ini}=45^\circ$, parallel and perpendicular component of the incident polarisation are equal and the resulting scattering polarisation is that of initially unpolarised radiation (cf. black solid line and red squares in Fig.~\ref{fig_ips3}). Larger incident polarisation angles ($\gamma_\text{ini}>45^\circ$) cause an amount of scattered polarisation that is higher than that for  initially unpolarised radiation. Incident polarisation angles larger than  $90^\circ$ lead to the contrary effect as the proportion of the incident polarisation perpendicular to $\vec{S}$ rises again. 

The differences between \mbox{$a_\text{eff}=\unit[100]{\mu m}$} and \mbox{$a_\text{eff}=\unit[150]{\mu m}$} grains are comparable to the differences when changing the incident polarisation degree (Section~\ref{41}). The scattering polarisation phase functions of the \mbox{$\unit[150]{\mu m}$} grains are more asymmetric and the peak polarisations are smaller, shifted to smaller scattering angles and decreasing with increasing polarisation angle $\gamma_\text{ini}$ of the incident radiation. 

To summarise, the highest amounts of linear polarisation by scattering on oblate dust grains can be reached when the incident polarisation is high and is directed parallel to the grain symmetry axis $\vec{S}$. As this effect increases with the axis ratio of the grains, scattering of intrinsic polarised radiation on non-spherical grains allows higher polarisation fractions than in the case of spherical grains. On the other hand, the incident polarisation can lead to a depolarisation for most of the scattering angles when the proportion of incident polarisation perpendicular to $\vec{S}$ is larger than the component parallel to it. Moreover, a 90~degree flip of the orientation of the scattering polarisation can occur at certain scattering angles.

\subsection{Circular polarisation}
\label{43}
Several observations of circumstellar environments have found high values of circular polarisation which might origin from dust scattering (\citealt{Whitney2002}). In Section~\ref{circ} we discussed the zero circular polarisation of scattering of unpolarised, incident radiation on oblate dust grains. \cite{Gledhill2000} showed that scattered light is significantly circularly polarised when the symmetry of the scattering event is broken. Besides a change of the orientation of the symmetry  axis or the scattering planes, this is realised when the incoming wave is initially linearly polarised. In order to break the symmetry, the incident polarisation orientation has to be misaligned with both the symmetry axis $\vec{S}$ and the scattering plane. Consequently, we find that the scattered radiation is circularly polarised when $\gamma_\text{ini}\neq 0^\circ$ or $\neq90^\circ$ (Fig.~\ref{fig_circ}). At $\gamma_\text{ini}=45^\circ$, the circular polarisation is maximum with absolute values of $P_\text{sca,circ}{\sim}\unit[6]{\%}$ ($a_\text{eff}=\unit[100]{\mu m}$; Fig.~\ref{fig_circ}, \textit{left}) and $\unit[{\sim}32]{\%}$ ($a_\text{eff}=\unit[150]{\mu m}$; Fig.~\ref{fig_circ}, \textit{right}). A flip from left- to right-handed circular polarisation occurs for all polarisation phase functions.

\section{Discussion}
\label{501}
We have seen in Sections~\ref{301} and \ref{401} that non-spherical grain shapes have a significant effect on the scattering and in particular the emerging scattering polarisation. While the deviations between spherical and oblate grains of sizes below $\unit[100]{\mu m}$ are mostly confined on forward and backwards scattering angles, larger grains with $a_\text{eff}>\unit[100]{\mu m}$ show significant deviations in scattering polarisation values,  polarisation  peaks and polarisation orientations when oblate grains are taken into account. This reveals clearly the different scattering behaviour of spherical and non-spherical grains once the grain sizes are beyond the Rayleigh~regime, i.e., in the Mie regime. Besides single-size grains, also grain size distributions with maximum sizes above $\unit[100]{\mu m}$ show a clear dependence on the particle shape. In the literature, the interpretation of polarisation observations is commonly based on a perfect spherical, compact dust model. Our results show that the grain sizes derived from self-scattering on spherical grains give the Rayleigh limit of the observation wavelength, however, not the actual grain size. Therefore, the grain sizes deduced from self-scattering polarisation as discussed in current literature have to be re-evaluated in order to avoid misleading conclusions. 

Scattering of non-spherical grains allows the production of higher polarisation fractions. As polarisation values larger than $\unit[\sim5]{\%}$ detected in observations have usually been interpreted as being produced by polarised emission of aligned grains, this rises questions about its origin and thus also on the inferred dust properties.

The scattering on non-spherical grains shows also significant circular polarisation. Contrary to spherical grains, single-scattering is sufficient to circularly polarise the scattered radiation (e.g.,~$\unit[33]{\%}$ for \mbox{$a_\text{eff}=\unit[150]{\mu m}$} grains) unless the grain symmetry axis is symmetric arranged to the polarisation orientation of the incoming wave. Therefore, upcoming observations of circular polarisation bear the potential to unveil the presence of non-spherical grains in protoplanetary disks.

We have seen further that an initial polarisation of the incoming wave can lead either to higher scattering polarisation degrees or to depolarisation, depending on the orientation of the non-spherical dust grain relative to the incident wave and on the scattering plane as well as on the incident polarisation degree and incident polarisation orientation. The incoming wave can be polarised due to intrinsic polarisation (polarisation by emission) of the non-spherical grains. For a wavelength of $\lambda=\unit[870]{\mu m}$ and an effective radius of \mbox{$a_\text{eff}=\unit[100]{\mu m}$}, the direction of the intrinsic polarisation is perpendicular to the symmetry axis $\vec{S}$ (\citealt{Kirchschlager2019}). If we assume perfect alignment of oblate grains in a protoplanetary disk (due to radiation or magnetic fields), this leads to the scenario shown in Fig.~\ref{fig_ips2}. 

The higher polarisation fractions compared to spherical grains will be visible in protoplanetary disks or star-forming regions when the dust grains are well aligned. However, due to non-perfect alignment of the grains or multiple scattering, the polarisation state of both the incoming and the scattered wave might deviate. A prediction whether oblate grains cause depolarisation or larger scattering polarisation degrees is ambitious and has to be moved to sophisticated radiative transfer simulations which include grain alignment processes (e.g., \citealt{Bertrang2017a,Bertrang2017b}). Moreover, the presence of different dust materials and grain shapes (including prolate grains) in a disk will potentially affect the outcome as well. 

The non-spherical dust grains will also emit polarised radiation which adds to the polarised signal of the scattering. For the convenience and clarity of the presentation, we focussed on the polarisation
by scattering and did not consider the superposition of scattering and emission which can have polarization efficiencies at comparable levels. Future investigations will have to include not only self-scattering by oblate grains with radiation transfer in complex magnetic field geometry, but also polarised emission by aligned grains. These calculations will have to be done in the Mie regime for self-scattering and for polarised emission (\citealt{Guillet2020}), and not in the Rayleigh regime as it is usually done.

\section{Conclusions}
\label{601} 
We have investigated the polarisation by single-scattering on non-spherical dust grains. The polarisation degree as well as the polarisation orientation of the incoming wave were modified in order to study different scenarios. Our main findings are:
\begin{itemize}
 \item The scattering polarisation of oblate dust grains significantly deviates from that of compact spheres, both for incoming polarised and unpolarised radiation. In the Mie regime (e.g. for grain sizes $\unit[{>}100]{\mu m}$ at wavelengths \mbox{$\lambda=\unit[870]{\mu m}$)} the deviations between spherical and oblate grains are tremendous and the usage of spherical grains when interpreting polarisation observations is deficient. 
 \item Considering non-spherical in self-scattering simulations has the potential to explain polarisation observations with grains significantly larger than ${\sim}\lambda/2\pi$. Thus, the grain sizes derived from self-scattering simulations using non-spherical, perfectly aligned grains can be larger than that from simulations using spherical grains.
 \item Self-scattering by oblate, aligned grains produces higher polarisation fractions compared to spheres. This calls into question if high polarisation values ($\unit[\gtrsim 5]{\%}$) observed in protoplanetary disks or star-forming regions are produced by polarised emission, as usually assumed, or by self-scattering.
 \item Scattering of intrinsically polarised waves can lead to either depolarisation or amplification of the scattering polarisation, including polarisation flips. The exact results depend strongly on the grain alignment processes in protoplanetary disks as well as on disk and dust properties.
 \item Circular polarisation is a promising method for the verification of the presence of non-spherical grains and to unveil further dust parameters.
\end{itemize}

\begin{acknowledgements}
We thank the anonymous referee for her/his constructive comments. FK was supported by European Research Council Grant SNDUST ERC-2015-AdG-694520. GHMB acknowledges funding from the European Research Council under the European Union's Horizon 2020 research and innovation programme (grant agreement No. 757957).
\end{acknowledgements}
 \bibliography{IPS}
 \bibliographystyle{aa}

\appendix
\section{Linear polarisation as a function of wavelength}
\label{waves}
 \begin{figure*}                                
\centering                                       
\includegraphics[trim=2.5cm 2.6cm 2.1cm 2.5cm,  clip=true, width=0.3333\linewidth, page = 3]{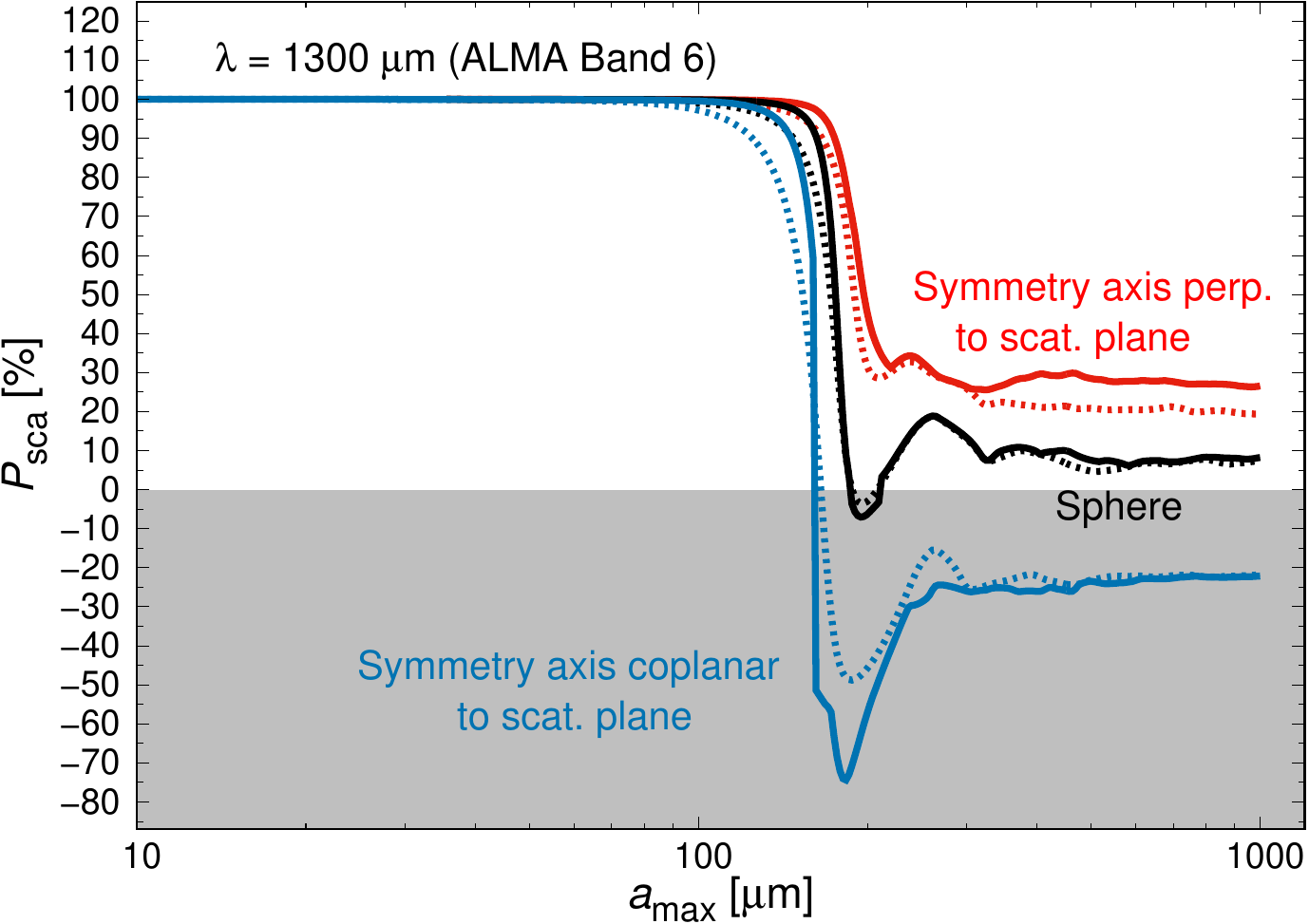}\hspace*{0.1cm}
\includegraphics[trim=2.5cm 2.6cm 2.1cm 2.5cm,  clip=true, width=0.3333\linewidth, page = 5]{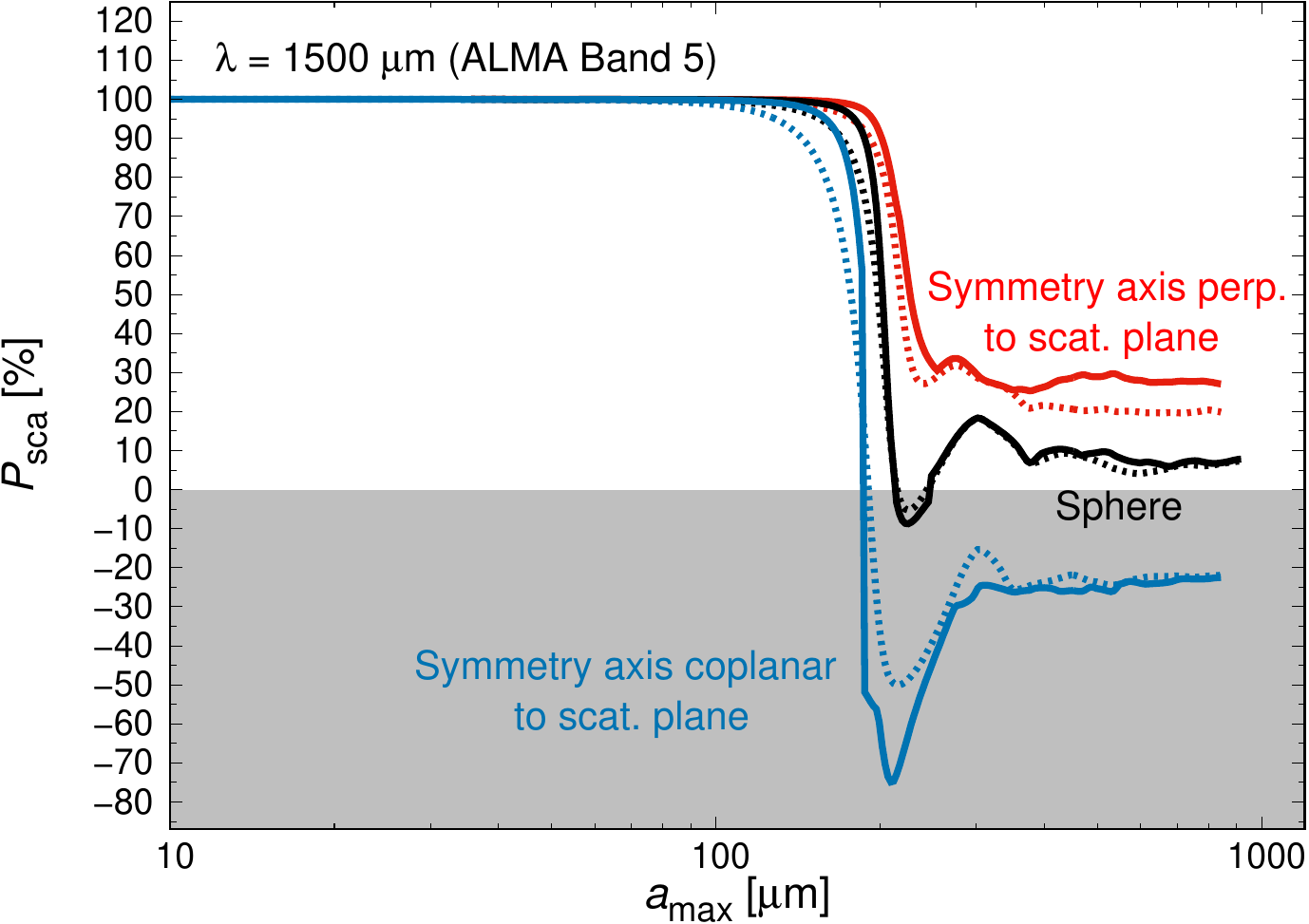}\hspace*{0.1cm}
\includegraphics[trim=2.5cm 2.6cm 2.1cm 2.5cm,  clip=true, width=0.3333\linewidth, page = 7]{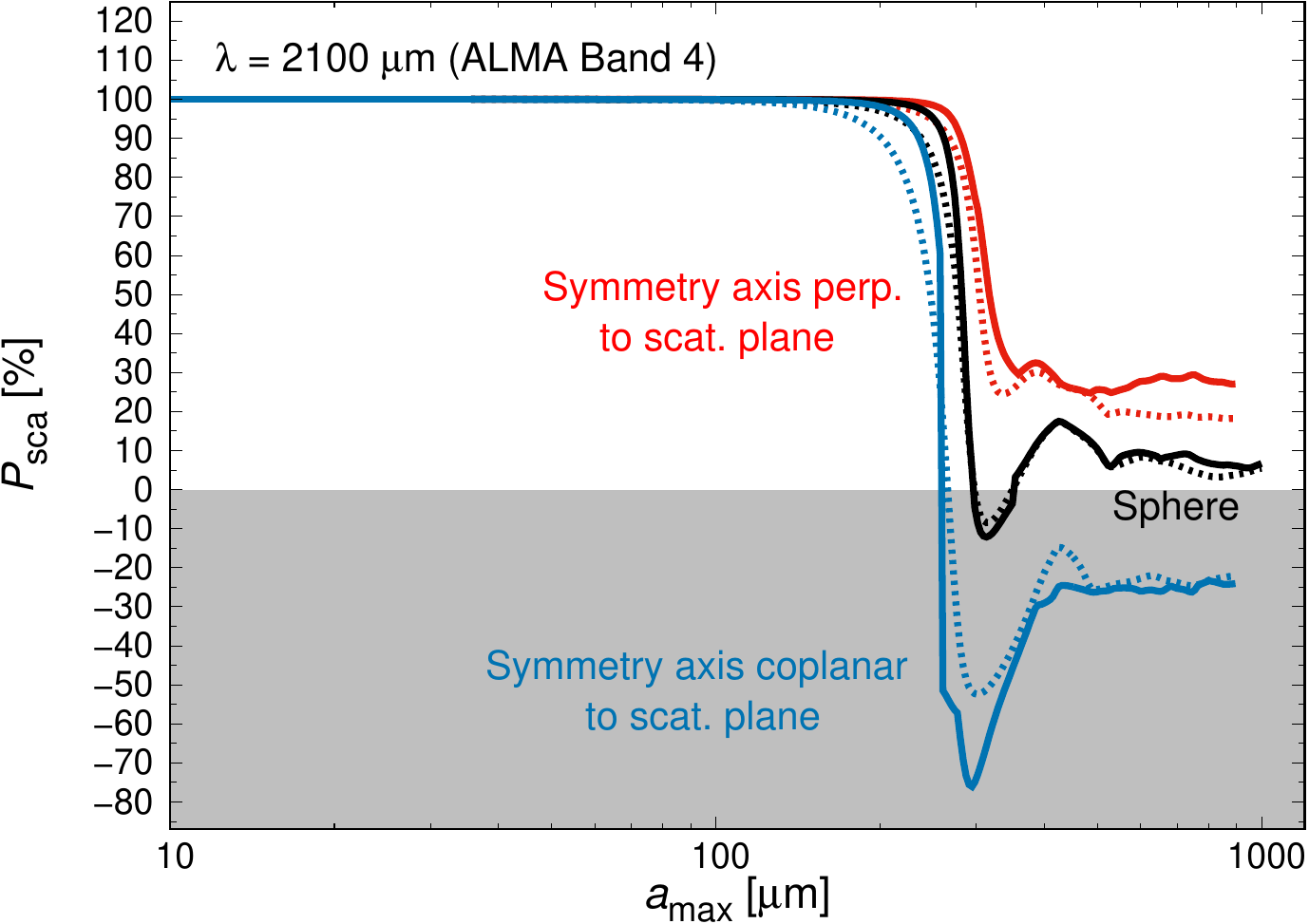}\\
\includegraphics[trim=2.5cm 2.6cm 2.1cm 2.5cm,  clip=true, width=0.3333\linewidth, page = 9]{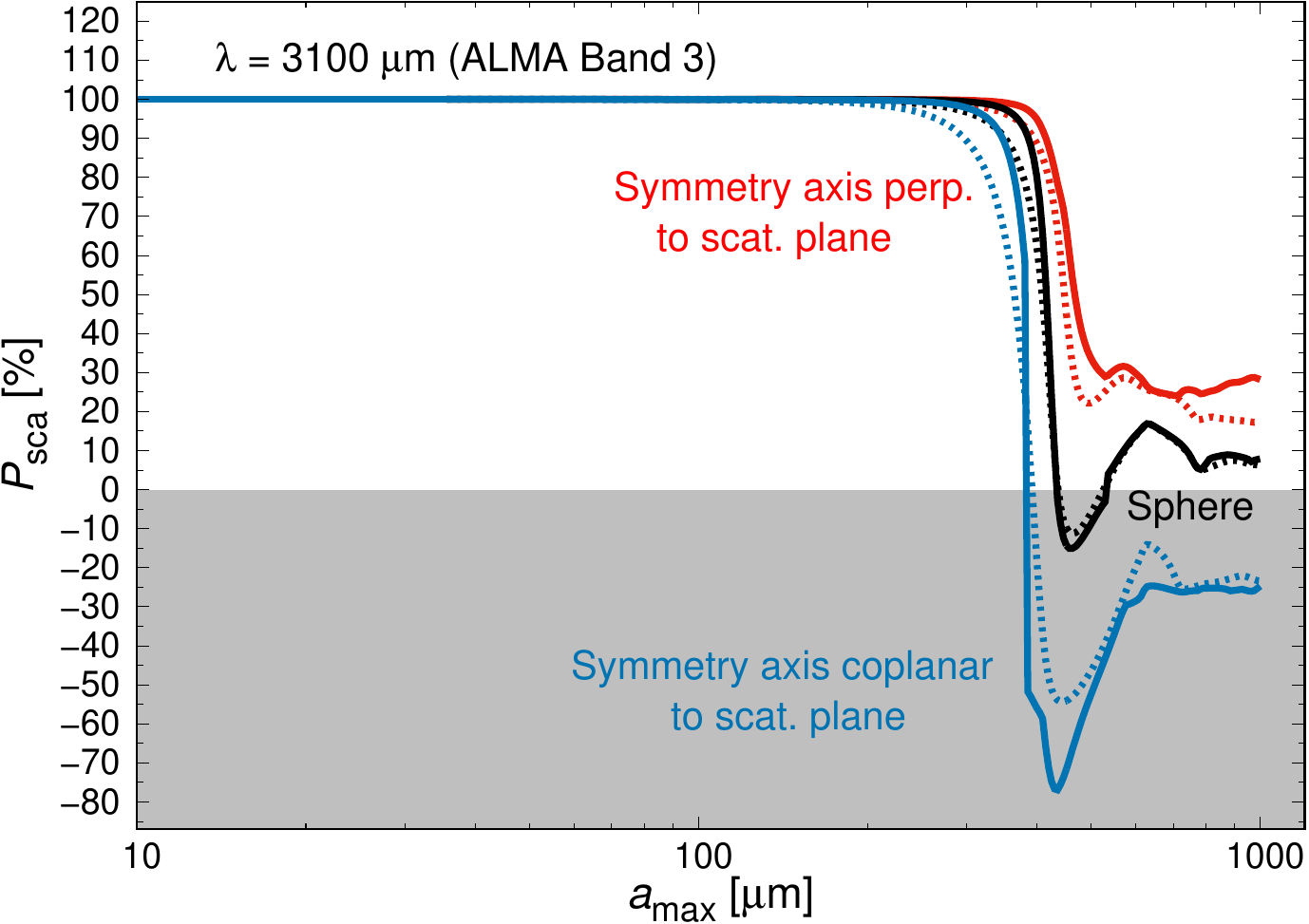}\hspace*{0.1cm}
\includegraphics[trim=2.5cm 2.6cm 2.1cm 2.5cm,  clip=true, width=0.3333\linewidth, page =11]{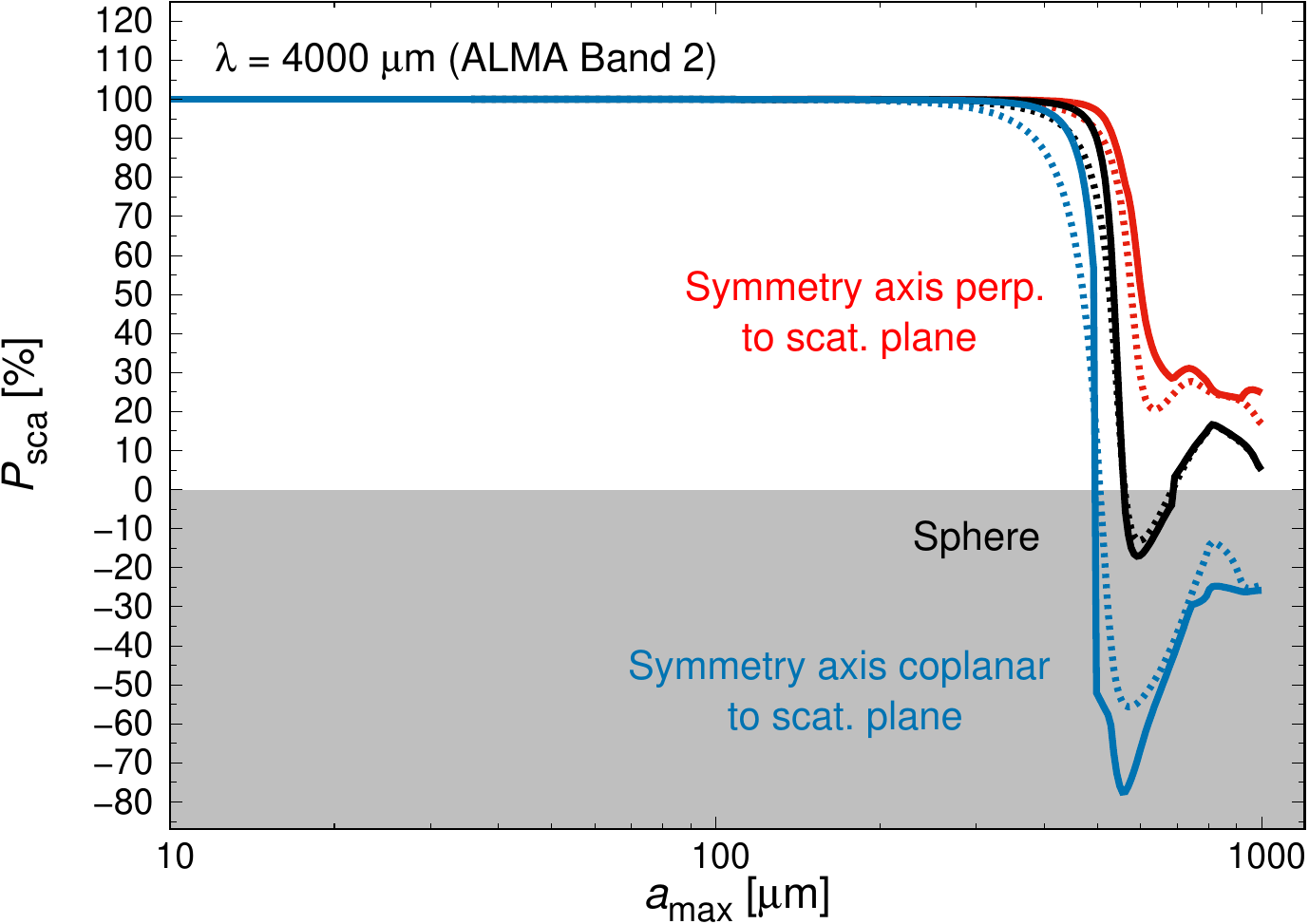}\hspace*{0.1cm}
\includegraphics[trim=2.5cm 2.6cm 2.1cm 2.5cm,  clip=true, width=0.3333\linewidth, page =13]{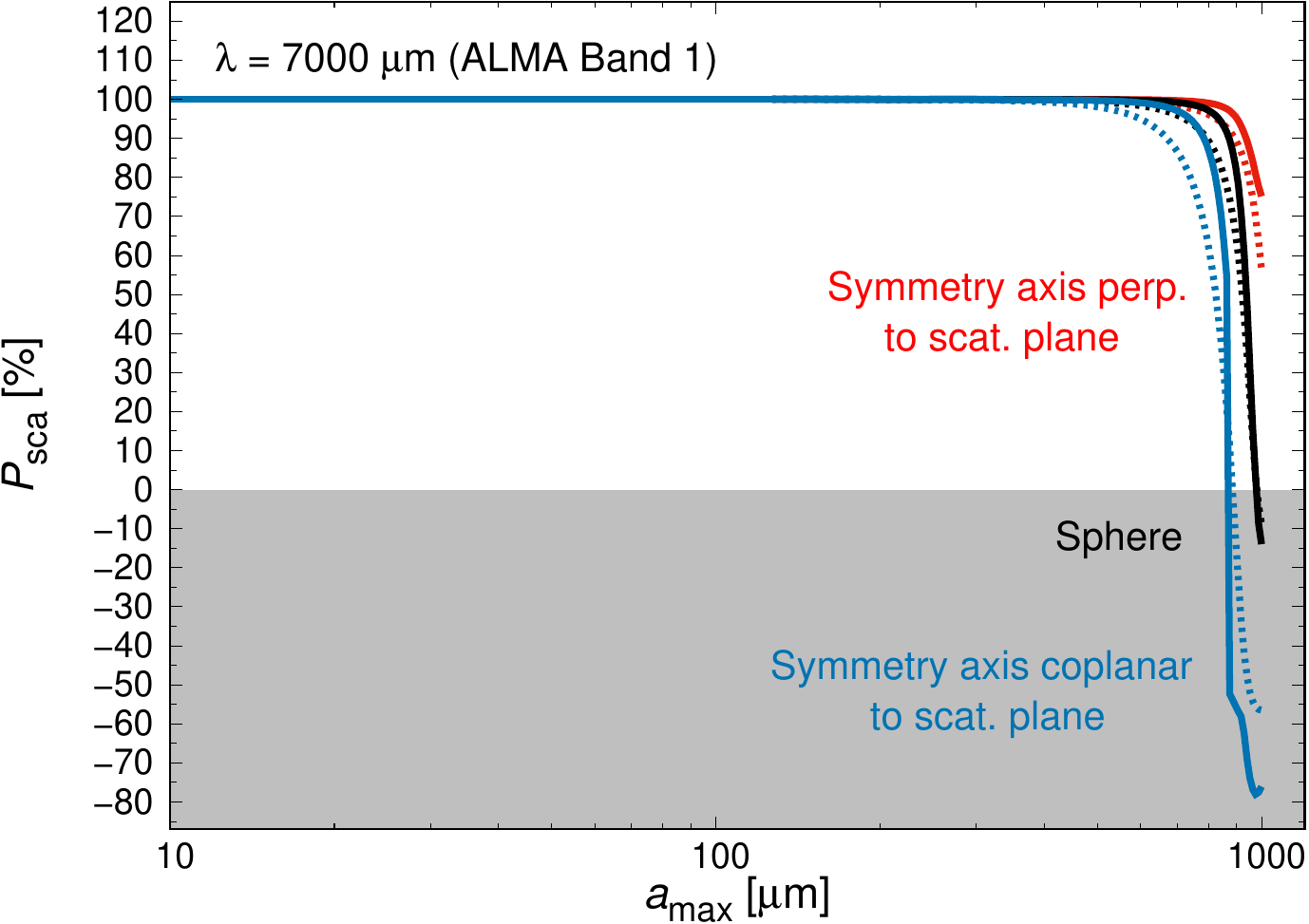}
\caption{Same as Fig.~\ref{fig_distrib}, only for the ALMA wavebands B6, B5, B4, B3, B2, and B1 ($\lambda=\unit[1.3]{mm},\unit[1.5]{mm},\unit[2.1]{mm}, \unit[3.1]{mm},\unit[4]{mm}, \unit[7]{mm}$, resp.).}                                      
\label{fig_waves}                             
\end{figure*} 
We calculated the scattering polarisation for grain size distribution  $\text{d}n\propto a_\text{eff}^{-3.5}\text{d}a_\text{eff}$ for the different ALMA wavebands B6, B5, B4, B3, B2, and B1 ($\lambda=\unit[1.3]{mm},\unit[1.5]{mm},\unit[2.1]{mm}, \unit[3.1]{mm},\break \unit[4]{mm}$ and $\unit[7]{mm}$, resp.). The minimum effective radius of the grain is $a_\text{min}=\unit[5]{nm}$ while the maximum grain size is varied between $\unit[10]{\mu m}$ and $\unit[1]{mm}$ (Fig.~\ref{fig_waves}). The Rayleigh regime is linearly shifted with wavelength $\lambda$ to larger maximum grain sizes while the scattering polarisation differences between spherical and non-spherical grains are preserved.

\section{Scattering formalism}
\label{Stokes}
The scattering properties of a dust grain are described by its $4\times 4$ scattering matrix which is a function of grain size, wavelength, material,  dust grain shape (e.g.,~elongicity) and morphology (e.g.,~porosity) as well as the scattering angles $\Theta_\text{sca}$ and $\Theta_\text{sca,2}$. Scattered radiation is characterised by the Stokes vector $(I_\text{sca},Q_\text{sca},U_\text{sca},V_\text{sca})$ which is the product of the scattering matrix  and the Stokes vector $(I_\text{ini},Q_\text{ini},U_\text{ini},V_\text{ini})$ of the incident wave. 
For oblate dust grains (in contrast to compact spheres), all scattering matrix elements can have a significant contribution. The Stokes vector of the scattered radiation is then given by 
\begin{align}
 \begin{pmatrix}
 I_\text{sca}\\
 Q_\text{sca}\\
 U_\text{sca}\\
 V_\text{sca}
 \end{pmatrix}
 =
 \begin{pmatrix}
 S_{11}&S_{12}&S_{13}&S_{14}\\
 S_{21}&S_{22}&S_{23}&S_{24}\\
 S_{31}&S_{32}&S_{33}&S_{34}\\
 S_{41}&S_{42}&S_{43}&S_{44}\\
 \end{pmatrix}
 \cdot
 \begin{pmatrix}
 I_\text{ini}\\
 Q_\text{ini}\\
 U_\text{ini}\\
 V_\text{ini}
 \end{pmatrix}
 \begin{matrix}
 \\
 \\
 \\
 .
 \end{matrix}
 \label{eqa1}
 \end{align}      
The linear polarisation degree and the polarisation orientation of the scattered/incident radiation are then
\begin{align}
\begin{split}
P_\text{sca/ini}&=\left(\frac{Q_\text{sca/ini}^2+U_\text{sca/ini}^2}{I_\text{sca/ini}^2}\right)^{0.5}\\
      \text{and}&\\
\gamma_\text{sca/ini}&=\, \frac{1}{2}\arctan{\left(\frac{U_\text{sca/ini}}{Q_\text{sca/ini}}\right)}+l\,\frac{\pi}{2}\,,
\end{split}
\end{align}
where $l=1$ if $Q_\text{sca/ini}<0$, and $l=0$ otherwise.

For a special orientation of the grain relative to the incoming wave (e.g., $\gamma_\text{axi}=90^\circ, \Theta_\text{sca,2}=0^\circ$ or $90^\circ$), equation~(\ref{eqa1}) can be simplified similar to spherical grains and scattered linear and circular polarisation are calculated  by equation~(\ref{eq01}) and (\ref{eq02}), respectively.

In order to realise a linearly polarised, incoming wave with polarisation degree $P_\text{ini}$ and polarisation orientation \mbox{$\gamma_\text{ini}=0^\circ,22.5^\circ,45^\circ, 67.5^\circ$,} and $90^\circ$, we set the Stokes vector of the incident radiation to $(1,P_\text{ini},0,0)$, $(1,\nicefrac{P_\text{ini}}{\sqrt{2}},\nicefrac{P_\text{ini}}{\sqrt{2}},0)$, $(1,0,P_\text{ini},0)$, $(1,\nicefrac{-P_\text{ini}}{\sqrt{2}},\nicefrac{P_\text{ini}}{\sqrt{2}},0)$, and $(1,-P_\text{ini},0,0)$, respectively.

\end{document}